\documentclass{article}
\usepackage[fontsize=12pt]{fontsize}
\usepackage{hyperref}       % hyperlinks
\hypersetup{colorlinks = true,  
	linkcolor = blue, 
	urlcolor = blue,%网页链接为蓝色
	citecolor = blue} %文献引用颜色设置为蓝色
\usepackage[verbose=true,letterpaper]{geometry}
\AtBeginDocument{
	\newgeometry{
		textheight=9in,
		textwidth=7in,
		top=1in,
		headheight=14pt,
		headsep=25pt,
		footskip=30pt
	}
}            
\usepackage{setspace}
\usepackage{graphicx}
\usepackage[numbers,square,super,comma,compress]{natbib}
\usepackage{booktabs}       % professional-quality tables
\usepackage{amsfonts}       % blackboard math symbols
\usepackage{amsmath}
\usepackage{tabularx}
\usepackage{multirow}
\usepackage{makecell}
\usepackage{float}
\usepackage{bm}     %加粗

\title{\textbf{Phase transition like behaviors of propagation of passenger stranding phenomena in subway networks}}

\author{Xinyi Li$^1$, Shengda Zhao$^1$, Liang Wang$^{2, 3}$, Qing Wang$^1$,\\ Xun Zhang$^4$, Fang Liu$^4$, Xiaodong Zhang$^2$,\\ Daqing Gong$^3$, Xinghua Zhang$^1$\thanks{Corresponding author.\\\texttt{21121588@bjtu.edu.cn} (Xinyi Li), \texttt{21121598@bjtu.edu.cn} (Shengda Zhao), \texttt{wljean@126.com} (Liang Wang), \texttt{21121592@bjtu.edu.cn} (Qing Wang), 
\texttt{zhangxun@jcmetro.com} (Xun Zhang),
\texttt{liufang@jcmetro.com} (Fang Liu)
\texttt{zhangxd-bicp@outlook.com} (Xiaodong Zhang), \texttt{dqgong@bjtu.edu.cn} (Daqing Gong), \texttt{zhangxh@bjtu.edu.cn} (Xinghua Zhang).}\\[1cm]	
	$^1$School of Physical Science and Engineering, Beijing Jiaotong University,\\ Beijing 100044, China	\\
	$^2$Beijing Municipal Institute of City Planning and Design,\\ Beijing 100045, China\\
	$^3$School of Economics and Management, Beijing Jiaotong University,\\ Beijing 100044, China\\
        $^4$Capital Metro Co.,Ltd, Beijing 101300, China\\[0.5cm]
}

\date{ }
\begin{document}
\maketitle 	
\setstretch{1.5}

\begin{abstract}
The subway as the most important transportation for daily urban commuting is a typical non-equilibrium complex system, composed of 2 types of basic units with service relationship. One challenge to operate it is passengers be stranded at stations, which arise from the spatiotemporal mismatch of supply scale and demand scale. More seriously, there is a special phenomenon of the propagation of passenger stranding (PPS) by forming stranded stations clusters, which significantly reduces the service efficiency. In this study, Beijing subway as an example is studied to reveal the nature of PPS phenomena from a view point of statistical physics. The simulation results demonstrate phase transition like behaviors depending on the ratio of service supply scale and demand scale. The transition point can quantitatively characterize the resilience failure threshold of service. The eigen microstate method is used to extracting the fundamental patterns of PPS phenomena. Moreover, this study offers a theoretical foundation for strategies to improve service, such as topological planning and train timetable optimization. The methodology developed in present work has significant implications for study of other service systems.
\end{abstract}

\section{Introduction}
Among various urban transportation modes, the travelling time by taking the subway is the most easily controlled due to its immunity from issues like road congestion and inclement weather. The carbon emissions per passenger transported and unit distance covered by electric-powered subways are significantly lower than other options. As a result, subways offer an optimal solution to address common urban issues such as traffic congestion, environmental pollution, and limited efficiency of public services in large cities. The spatial extension characteristics of subways also shape the structure of large cities and the characteristics of residents' activities, known as Transit-Oriented Development (TOD)\cite{1}. By taking advantage of the economy and efficiency of the subway, residents can opt to reside in low-cost suburbs while pursuing employment opportunities in the city center, resulting in a significant spatial decoupling between job and residence centers. As the most crucial public service system in cities, the subway system bears the responsibility to provide efficient services to meet the highly concentrated daily commuting demands. Its efficiency is primarily characterized by its ability to fulfill the transportation demands of passengers while effectively reducing waiting time.

\begin{figure}[!ht]
	\centering
	\includegraphics[width=0.82\textwidth]{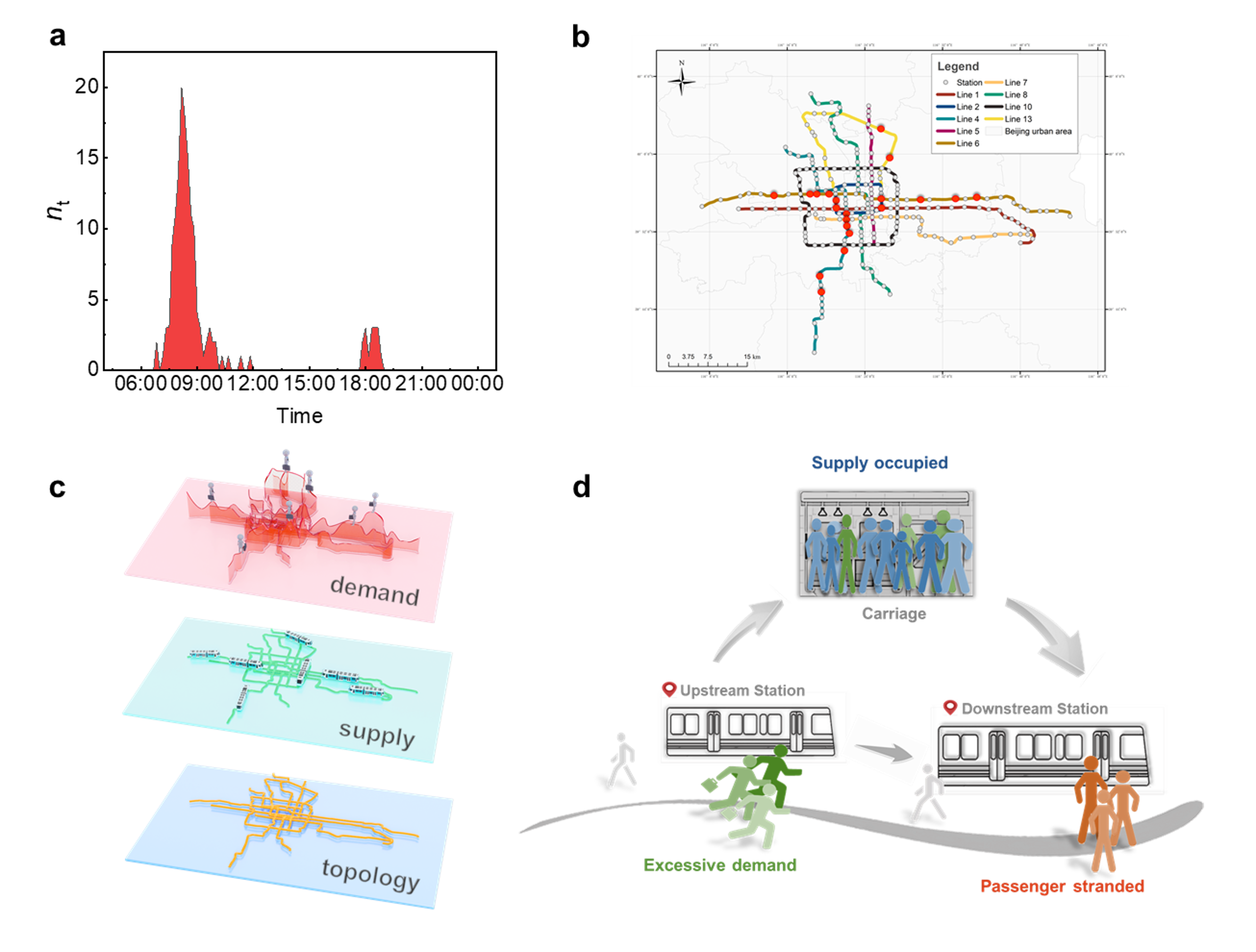}
	\caption{Stranding phenomena is a major problem of subway system: \textbf{a} The temporal variation in the number of stations with stranded passengers in the subway network throughout a day on May 10, 2021. \textbf{b} Spatial distribution of stations with stranded passengers in the Beijing subway network at 8:10 AM rush hour. \textbf{c} Mechanisms of passengers being stranded at subway stations. The causes for the occurrence of stranded passengers and PPS phenomena are as follows: network topology, service trains scheduling, and spatiotemporal characteristics of demand. \textbf{d} the scheme of causing of PPS phenomena. stranding occurs at stations when passengers’ demand exceeds transportation supply. Due to the interconnected nature of the line, operation in full capacity in upstream stations would occupy the supply scale of downstream stations, leading to passengers stranded at the latter.}
	\label{fig:fig1}
\end{figure}

Because of the discrete arrival of trains as service providers, coupled with the randomness of passenger arrivals as service demanders, there are invariably random incidents of passengers being stranded in stations. Generally, there is no temporal correlation between these events at stations. When the passenger arrival rate at a station exceeds the efficiency of the subway's service, the events of passenger stranding start to exhibit temporal correlation. During the morning rush hours, the Beijing subway system encounters the most severe instances of stranding, with 8.547\% of stations affected. Fig.~\ref{fig:fig1}a illustrates the number of stations with stranding phenomena over time on Monday, May 10, 2021. The stranding phenomenon can propagate from station to station along the subway line, and these stranded stations form a spatially correlated cluster. We call this phenomenon the propagation of stranded passengers (PPS). As an example, the stranded stations during the morning rush hours are illustrated in Fig.~\ref{fig:fig1}b. The presence of temporal and spatial correlations of stranding in the subway not only leads to longer commuting times for passengers, but also undermines the advantages of subway travel and depresses societal efficiency. Large-scale passenger stranding within enclosed subway stations increases the difficulty of evacuation and raises concerns regarding public safety, such as the spread of epidemics and the risk of terrorist attacks. 

When the duration and spatial scale of stranding reach a substantial magnitude, the system is unable to provide adequate service supply to meet the demand. The critical demand scale in this case can be defined as the threshold of service resilience, which serves as a crucial parameter for quantitative evaluation in decision-making processes related to investment, insurance, risk management, and emergency preparedness. The term ``resilience" originated in mechanics to describe the resistance of materials to physical impacts. In the study of complex systems, the concept of resilience can be primarily categorized into 3 frameworks\cite{2}: (1) System resilience (or ecological resilience) focuses on the magnitude of the change or perturbation that a system can endure without shifting to another stable state\cite{3}; (2) Engineering resilience is defined by the recovery rate or time\cite{4}; (3) Adaptive resilience characterizes the capacity of socioecological systems to adapt or transform in response to unfamiliar, unexpected and extreme shocks\cite{5}. The resilience of services studied here can be attributed to the category of system resilience. The system's responsive properties, as characterized by the stranding scale, are measured by adjusting the demand scale, which acts as the stimulus. The threshold of resilience is defined as the demand scale at which a notable change in response behavior occurs, beyond which the service efficiency of the system is significantly decreased.

Generally, the subway has been studied using traffic fluid models\cite{6} or complex networks\cite{6}. The traffic fluid models are inspired by the studies of congestion in road traffic\cite{7,8,9,10,11,12}. The fundamental diagram of traffic flow and velocity reflects the spontaneous phase transition of congestion\cite{5}. Subway studies based on complex networks investigate the reliability of efficient travel paths by establishing undirected network models\cite{13,14} or directed network models\cite{15}, considering the network connectivity characteristics of stations. Unlike traffic flow in roads and transportation problems in complex networks, the PPS phenomenon stem from the timely fulfillment of service relationships locally at each station governed by the ratio of service supply rate and the demand rate, as well as the factors of topological of network as shown in Fig.~\ref{fig:fig1}c. The service supply rate is determined by the passenger capacity of trains and the arrival rate of trains (timetable), and the demand rate is determined by the arrival rate of passengers at this station. The development patterns of TOD lead to highly spatiotemporally uneven demands for subway services due to the tidal characteristics of commuting during morning and evening rush hours\cite{16,17}. This makes it challenging to achieve spatial and temporal matching between service supply and demand. The topological factor depends on both the connections of the stations to the network and the lines along which the trains provide service. This is because the passengers cannot travel freely within the networks, but must rely on trains to travel along predefined lines. If the planned route of a demand needs to pass through 2 or more subway lines, transferring to different trains is needed. Besides the local mismatching of demand and supply, it is worth noting that the stranding conditions at upstream stations would reduce the supply rate of downstream station as well due to the characteristics of the train providing service along the lines, as depicted in Fig.~\ref{fig:fig1}d. When the train operates at almost full capacity, the number of passengers getting off the train might be lower than those who want to get on, which gives rise to the propagation of stranding phenomena along the line. The TOD mode contributes to a significant spatial separation between job and residence centers, resulting in a large average commuting distance for passengers, which determines the spatial correlation length of PPS phenomena. 

The transportation in subway system is a collective behavior of a vast number of active basic units (passengers and trains), making it a typical subject of study in on-equilibrium statistical physics\cite{18}. Different from the systems composed of a single type of unit, the subway system is a service system consisting of 2 distinct types of units interplay through service relationship: (1) passengers as the demander of services and (2) trains as the supplier of services. A demand of service from a single passenger is a pair of stations depending on time called as origin-destination (OD) pairs, and on the supply side of services, subway trains operate according to the schedules of different lines. 
The formation of clusters of stranded stations, which serves as the main characteristic of PPS\cite{19}, exhibits similarities to phase transition phenomena similar to the condensation of gas molecules attracting and coalescing into a liquid state during vapor-liquid transitions, and the growth process of magnetization in ferromagnetic materials. To reveal the mechanism underlying the PPS phenomenon, it is essential to conduct research employing statistical physics theories and methodologies.

In order to determine the service resilience threshold, it is necessary to investigate the response behavior of subway systems under varying demand scales. However, executing controlled variable experiments within the actual infrastructure of large cities, particularly when aiming to surpass the resilience threshold, encounters substantial obstacles. As a typical non-equilibrium system, a data-driven dynamic model can be established, enabling the acquisition of its response patterns under different demand scales through computer simulations. In turn, the service resilience threshold can be predicted. In the present work, the system's responsive behaviors which are characterized by both the scale of stranding and the average waiting time of passengers are investigated in both steady-state and non-steady-state non-equilibrium conditions. Moreover, the PPS phenomenon encompasses a multitude of coupling effects within the subway system. Extracting the fundamental patterns from these coupling effects can provide valuable insights for the development of subway infrastructure, regulation of social services, and the establishment of effective methods for emergency response and incident prediction. In the field of non-equilibrium statistical physics, eigen microstate method offers a powerful method to investigate phase transitions, critical phenomena, and dynamic evolutions in complex systems, thereby providing a powerful tool for studying PPS phenomena. By applying the eigen microstates method\cite{20,21,22,23,24,25}, fundamental patterns and their evolutionary rules can be extracted from system evolution data. In present work, PPS as a class of emergent behaviors in self-organized multi-body systems are studied from a statistical physics perspective, using eigen microstates method to uncover the underlying mechanisms.

The service system can be considered as a general model for many economic and social activities of human beings. In these systems the scheduling of supply is optimized to meet the spatiotemporal characteristics of demand and PPS phenomena is a common feature of these systems. Furthermore, global operational efficiency, cost-effectiveness and preventing system collapse are essential considerations from a societal management point of view. Examples of service systems include transportation systems, like subway, taxis\cite{26}, buses\cite{27,28} etc; elevators in skyscrapers\cite{29,30,31,32}; medical appointment systems; logistical management\cite{33,34}; aircraft carriers decks management\cite{35,36,37}. In present work, the subway system as a reprehensive example of service system is studied and the mechanism of PPS phenomena of service system is revealed. The results and methodologies may provide valuable insights for service systems.

\section{Results}
To investigate the phenomenon of PPS in subway systems, a dynamic model based on an actual subway network is developed, using Beijing as a case study. Beijing is a world-class megacity with a population of 23 million permanent residents. As of July 26, 2022, the Beijing subway network is composed of 17 lines with a total of 330 operational stations and 69 transfer stations, constituting a transportation system of significant scale. The job and residence centers in Beijing exhibit characteristic features of TOD, with a substantial portion of the population relying on the subway for their daily commuting demands. The model system used in the present work, as shown in Fig.~\ref{fig:fig1}b, includes 3 east-west linear lines (Lines 1, 6, and 7), 3 north-south linear lines (Lines 4, 5, and 8), and 3 circular lines (Lines 2, 10, and 13), resulting in a subway network comprising 234 stations ($N = 234$). Among these stations, there are 36 transfer stations and 10 terminal stations (ends), capturing the representative topological structure of the Beijing subway system. These 9 earliest established lines, operating in the core area of Beijing serve as the main arteries in the real subway system. This model system is an appropriate representation in terms of scale, significance, and topological structure, which enables us to capture the daily operational features. In order to concentrate on the primary physical properties and alleviate computational burden, we have excluded the newly constructed lines from our study that have not yet achieved supply-demand equilibrium.

A dynamic evolutionary model system is proposed within this subway model to quantitatively investigate the phenomenon of PPS. This model system comprises 2 types of active basic units, passengers ($N_\mathrm{p} = 3,860,008$) and trains ($N_\mathrm{ori} = 5,266$) with service relationship. The service process includes the operation of trains to meet the demand of passengers. Trains operate according to a predetermined schedule, ensuring punctuality. When a train arrives at a station, passengers on board make decisions to disembark or not according to their destination demand, while passengers waiting at the station decide whether to get onboard based on the crowdedness of the compartment. Through computer simulations, the dynamic trajectories of passengers and trains can be obtained (further details can be found in Supplementary Information, A.2). To elucidate the fundamental physical characteristics of PPS, 2 simulation systems are considered: an ideal model (System \uppercase\expandafter{\romannumeral1}) and a realistic model (System \uppercase\expandafter{\romannumeral2}). In System \uppercase\expandafter{\romannumeral1}, it is assumed that the rates of demand and supply are independent of time and position, and the positions of OD pairs are randomly distributed throughout the network. This model system allows us to study PPS under non-equilibrium steady-state conditions. In System II\uppercase\expandafter{\romannumeral2}, the rate of passenger demand depends on  time, exhibiting with rush hour patterns. The positions of OD pairs capture the tidal nature of real commuting scenarios. The simulation time is measured in seconds, with the system state recorded every 5 minutes for System \uppercase\expandafter{\romannumeral1} and every 10 minutes for System \uppercase\expandafter{\romannumeral2}.

This study employs the average waiting time of passengers and the scale of stranding as the order parameters characterizing the PPS phenomena. The waiting time of passengers, $t_\mathrm{w}$, will increase due to the PPS phenomena, which serves as a crucial metric for evaluating the system's service efficiency. To quantify the PPS phenomena in the subway network, we develop here a quantitative framework based on a physical concept of cluster, which combines evolving traffic dynamics with network structure. Instead of the commonly used structural topology, only stations in the network with stranded passengers are considered functionally connected. In this way, we can characterize and understand the dynamics of PPS phenomena through cluster formation and vanishing. The scale of stranding in the system is characterized by 3 order parameters: the total number of stations within the stranding clusters,  $n_\mathrm{s}$; the size of the largest cluster, $m_\mathrm{c}$, represented by the number of stations within it; and the total number of stranded stations in the system, $n_\mathrm{t}$ (quantitative definition of the cluster can be found in Supplementary Information, B.1).

\subsection{System I: Steady-state features}
\begin{figure}[!ht]
	\centering
	\includegraphics[width=0.65\textwidth]{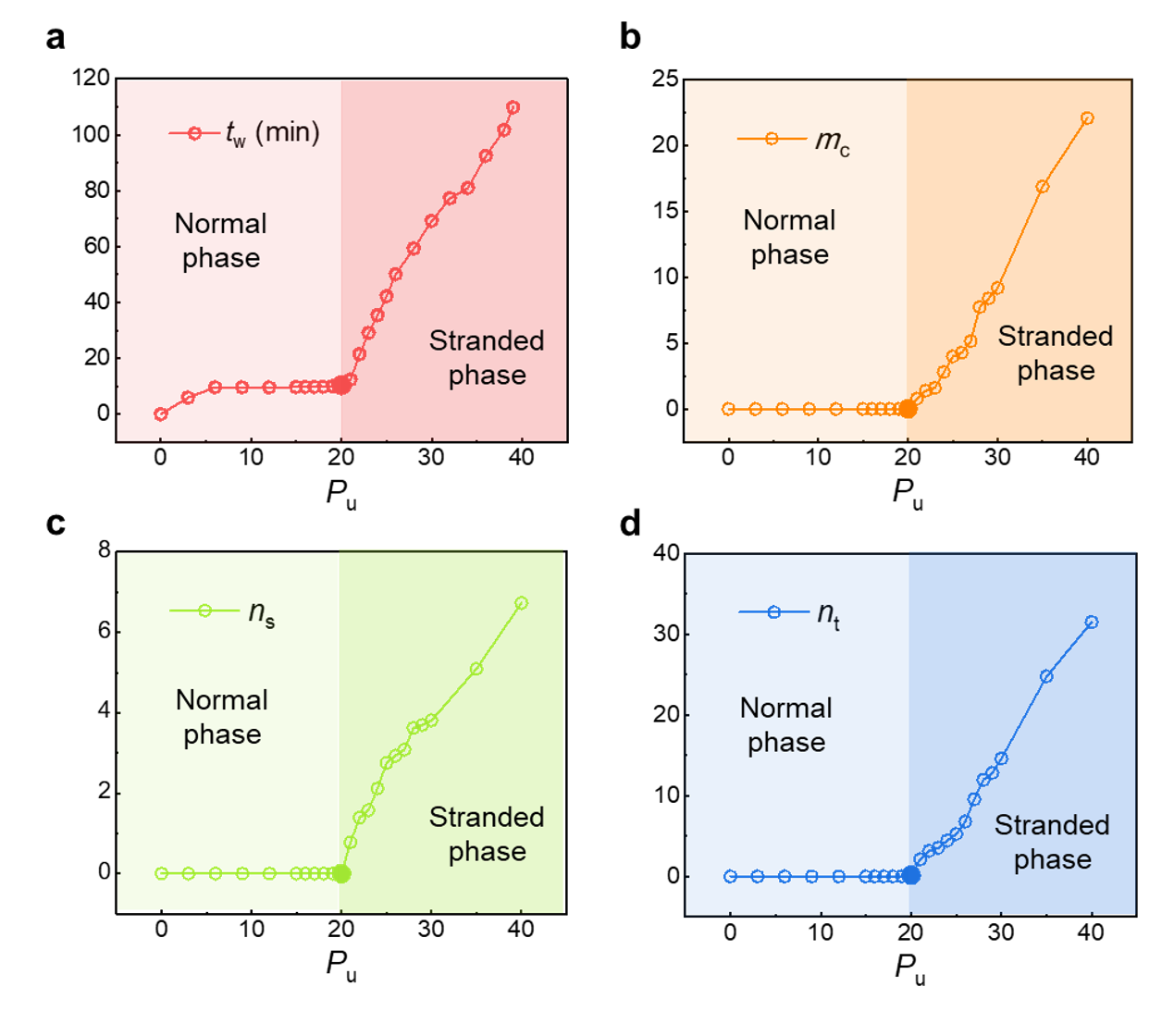}
	\caption{\textbf{a} Average waiting time of passengers, $t_\mathrm{w}$, as a responsive behavior of the demand scale, $P_\mathrm{u}$. \textbf{b}-\textbf{d} show the curves illustrating the changes in $n_\mathrm{s}$, $m_\mathrm{c}$, and $n_\mathrm{t}$ as responsive behaviors of $P_\mathrm{u}$. A phase transition like behavior occurs at $P_\mathrm{u} = 20$.
	}
	\label{fig:fig2}
\end{figure}

Assuming spatial and temporal homogeneity in both supply and demand, the system can reach a non-equilibrium steady state. By studying systems in this steady state, the influence of the subway network's topological structure on the PPS phenomenon can be revealed. In System I, supply scale remains constant by establishing a consistent train departure interval (4 min) from the originating station. The supply-demand ratio in the system by adjusting the demand scale, $P_\mathrm{u}$ (the number of passengers entering the station per minute). The specific simulation methodology can be found in the Supplementary Information, A.2.3. After a relaxation period of finite duration, the system reaches a non-equilibrium steady state, where the average waiting time of passengers and the values of these 3 order parameters steadied at specific values (examples in Supplementary Information, B.2.1, Fig.~B.4). The time dependent number of stranded passengers can be seen in Supplementary Information, B.2.2, Fig.~B.5, Fig.~B.6. System responsive behavior to the demand scale, $P_\mathrm{u}$, can be characterized by the waiting time $t_\mathrm{w}$, $n_\mathrm{s}$, $m_\mathrm{c}$ and $n_\mathrm{t}$ that are averaged with respect to time within steady state, which is shown in Fig.~\ref{fig:fig2}. Both $t_\mathrm{w}$ and the 3 order parameters exhibit similar responsive behaviors with a transition point around $P_\mathrm{u} = 20$ where the slopes of these 4 lines in Fig.~\ref{fig:fig2} show observable changes. When $P_\mathrm{u}$ is small, there are few stations with stranded passengers on the platforms randomly, and passengers can board the trains after a short waiting time. There is no spatial-temporal correlation between stations with stranded passengers, indicating the absence of the PPS phenomenon. Therefore, all 3 order parameters that characterize stranding scale tend to zero at small $P_\mathrm{u}$ values. When $P_\mathrm{u} > $20, $t_\mathrm{w}$ increases abruptly with $P_\mathrm{u}$, indicating a significant decrease in system efficiency. Stations with stranded passengers start to correlate with each other, forming clusters. The stranding scales indicated by all these 3 order parameters rapidly increase with increasing $P_\mathrm{u}$. The behavior that significant changes in its response properties beyond a certain threshold is similar to that of a continuous phase transition. The threshold value, $P_\mathrm{u}$ = 20, can be regarded as the transition point of this phase transition like behavior. When $P_\mathrm{u}$ is below this threshold, there is no correlation in the occurrence of stranding phenomena between the stations. We refer to this state as the normal phase. When $P_\mathrm{u}$ exceeds this threshold, stranded stations correlate with each other, forming persistent clusters of stations with stranded passengers that are difficult to dissipate. We refer to this state as the stranded phase. In this phase, the system can no longer provide normal service. Therefore, We can use the value of $P_\mathrm{u}$ at transition point as the threshold of resilience failure.

\begin{figure}[!ht]
	\centering
	\includegraphics[width=0.55\textwidth]{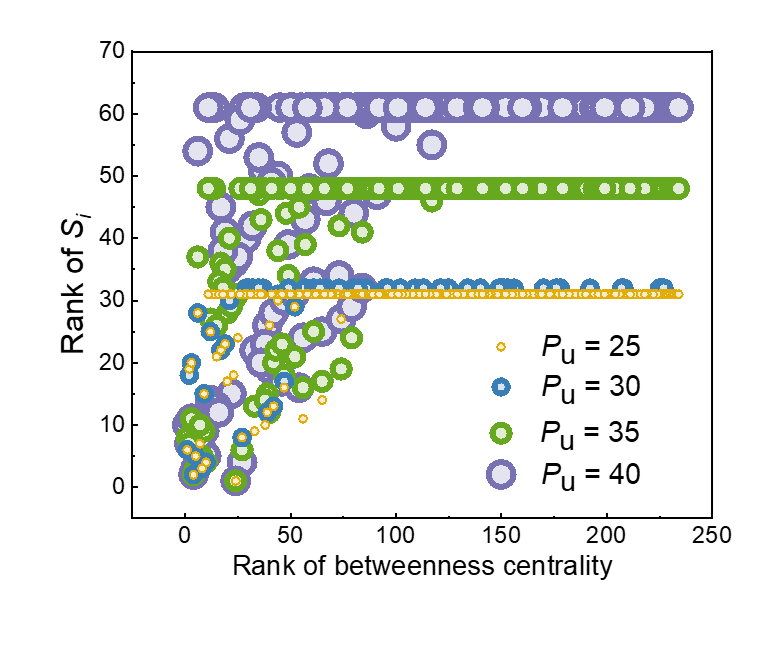}
	\caption{Relationship between the ranking of station betweenness centrality and the ranking of stranded passenger numbers which shows a positive correlation. Both 2 rankings are obtained through sorting in descending order based on the betweenness centrality values and the passenger number for each station.}
	\label{fig:fig3}
\end{figure}

Because there is no spatiotemporal matching issue between demand and supply in steady-state conditions, the formation of stranding and taking place of PPS phenomena solely arise from the topological factor. The topological factor is characterized by the network betweenness centrality of each station. We find that when system is in the stranded phase, most of the stranded stations tend to cluster around the stations with high network betweenness centrality. To depict the dependence between the occurrence of stranding at stations on the topological factor, we sort the stations in stranded phase (taking $P_\mathrm{u}$ = 25, 30, 35, and 40 as examples) in descending order based on the number of stranded passengers and the betweenness centrality, respectively. Using the ranks of these 2 queues as the vertical and horizontal coordinates in Fig.~\ref{fig:fig3}. If all stations fall on the diagonal line of this graph, the stranding phenomena and topological factors are linearly correlated. As indicated by Fig.~\ref{fig:fig3}, when systems are in the stranded phase, the stations with higher betweenness centrality exhibit relatively higher numbers of stranded passengers and those stations with lower betweenness centrality have no passenger stranded. Taking $P_\mathrm{u}$ = 30 as an example (blue circle), there are 33 stranded stations, with the maximum value of the vertical axis representing the ranking of stranded passengers being 34. Those stations with rank above 74 which have lower betweenness centrality do not experience passenger stranding.\\
\hspace*{2em}Generally, a station with higher betweenness centrality indicates a higher level of connectivity with the network, which would reduce the occurrence of stranding. However, results in Fig.~\ref{fig:fig3} show contrary to intuition. Indeed, this outcome reflects the characteristics of subway service. In systems under steady-state condition, since the passenger demands (pairs OD stations) are randomly generated in space, it has a high probability to require transfers to different lines at stations with higher betweenness centrality. Although these stations have high connectivity, passengers have to wait for the transfer train to arrive before they can continue their travel, resulting in a delay in receiving service. It should be noted that train scheduling in different lines is independent of each other, so passengers needing transfers generally cannot receive immediate service. This is equivalent to increasing the demand volume at the transfer stations acting as hubs, leading to stranding. By analyzing the steady-state conditions, the emergence of stranding phenomena and formation of clusters in the network are observed, which shows a phase transition-like behavior. The mechanisms underlying this behavior include the subway network's topology and the independent scheduling of different subway lines. The steady-state conditions analyzing is used to reveal the feature of spontaneous emergence of stranded clusters in the subway. However, such phenomena are not severe in real subway service, as the OD pairs of actual commuting passengers are not random. Citizens choose residence and workplace to minimize the possibility of transfers and enhance travel convenience. Hereafter, to investigate the PPS phenomenon in real systems, we will establish a computational model based on Automatic Fare Collection (AFC) data which provides the real OD pairs depending on time.

\subsection{System II: Spatiotemporal imbalance between supply and demand}

Large cities exhibit a large spatial separation between job and residence centers, particularly influenced by the TOD urban development model. Citizens tend to select residence and job locations along subway lines, resulting in a spatiotemporal imbalance in commuting demands with rush hours and long-distance travel patterns. To model the non-steady-state evolution of PPS, a tidal supply-demand model at a daily time scale based on the spatiotemporal characteristics of AFC data and train timetable data is constructed. The AFC data covers the entire weekdays from May 10th to May 14th, 2021, for the 9 subway lines depicted in Fig.~\ref{fig:fig1}b of the model (more details of passengers' related data in Supplementary Information, A.1.1). The AFC data consists of 3,860,008 records of passengers ($N_\mathrm{r}$) and 5,266 records of trains ($N_\mathrm{ori}$). The timetable data is from Beijing Rail Transit App (more details in Supplementary Information, A.1.2). The spatiotemporal mismatch between supply and demand is the primary mechanism of PPS phenomena. In the simulation, the supply scale is fixed at the real supply level while varying the demand scale to investigate the effects of service shortages on PPS. The demand scale is defined as $P_\mathrm{r} = N_\mathrm{d} / N_\mathrm{r}$, where $N_\mathrm{d}$ represents the total number of passengers taking subway throughout the day in the simulation, and $N_\mathrm{r}$ represents the real total number of passengers. We generate $N_\mathrm{d}$ OD pairs of demand by sampling the OD pairs in the same time interval based on the probability $P_\mathrm{r} = N_\mathrm{d} / N_\mathrm{r}$ from the real data throughout the weekdays (see Supplementary Information, A.2.4 and Fig.~A.3 for parameter settings and validation).

Unlike the results obtained under steady-state conditions, the computational results based on real-system data exhibit spatiotemporal heterogeneity. The time dependent number of stranded passengers can be observed in Supplementary Movie 1 for $P_\mathrm{r}=1.0$, and Supplementary Movie 2 for $P_\mathrm{r}=1.6$. In spatial dimension, there are 4 significant spatially independent clusters of stranded stations (Cluster \uppercase\expandafter{\romannumeral1}-\uppercase\expandafter{\romannumeral4}), regardless of the demand scale, whether it is low ($P_\mathrm{r} = 1.0$) (Supplementary Information, B.3.3, Fig.~B.8e) or high ($P_\mathrm{r} = 1.6$) (Supplementary Information, B.3.3, Fig.~B.8j). These clusters represent distinct commuting ``corridors" connecting the residence centers located in the suburb (north: Huilongguan, Tiantongyuan; south: Daxing; east: Beijing Municipal Administrative Center and North Three Counties, including: Sanhe City, Dachang Hui Autonomous County and Xianghe County; west: Shijingshan) to the job centers situated in the city center. The formation of these commuting corridors is attributed to the mutual influence of subway construction on the spatial distribution of residential and job areas, thereby characterizing Beijing as a TOD type city. The list containing the stations (names and labels) included in each cluster can be found in Supplementary Information, B.3.3, Table B.6 and B.7. In temporal dimension, instead of steady-state features, the system exhibits characteristics of rush hours. Several time points are selected to observe PPS, as depicted in Supplementary Information, B.3.3, Fig.~B.8. For a demand scale of $P_\mathrm{r} = 1.0$, at 7:10, only a few isolated stations manifest stranding phenomena, and these stations are spatially independent. By 8:00, PPS occurs, signifying the association of stations with stranded passengers and the formation of clusters. By 8:20, the size of these clusters reaches their maximum. The stranding phenomena completely vanish from the network by 13:00 (Supplementary Information, B.3.3, Fig.~B.8a-d). When the demand scale is high, such as $P_\mathrm{r} = 1.6$ (Supplementary Information, B.3.3, Fig.~B.8f-i), with the arrival of the morning rush hour, the system demonstrates associations between stations with stranded passengers, leading to cluster formation. As the morning rush hour subsides, the scale of PPS gradually decreased. In contrast to the case with a lower demand scale, during the onset of the evening rush hour at 13:00, the clusters generated during the morning rush hour do not entirely dissipate (Supplementary Information, B.3.3, Fig.~B.8i). The system is always in the working state with high stranding phenomenon, and the efficiency of service delivery is quite low. 

\begin{figure}[!ht]
	\centering
	\includegraphics[width=0.8\textwidth]{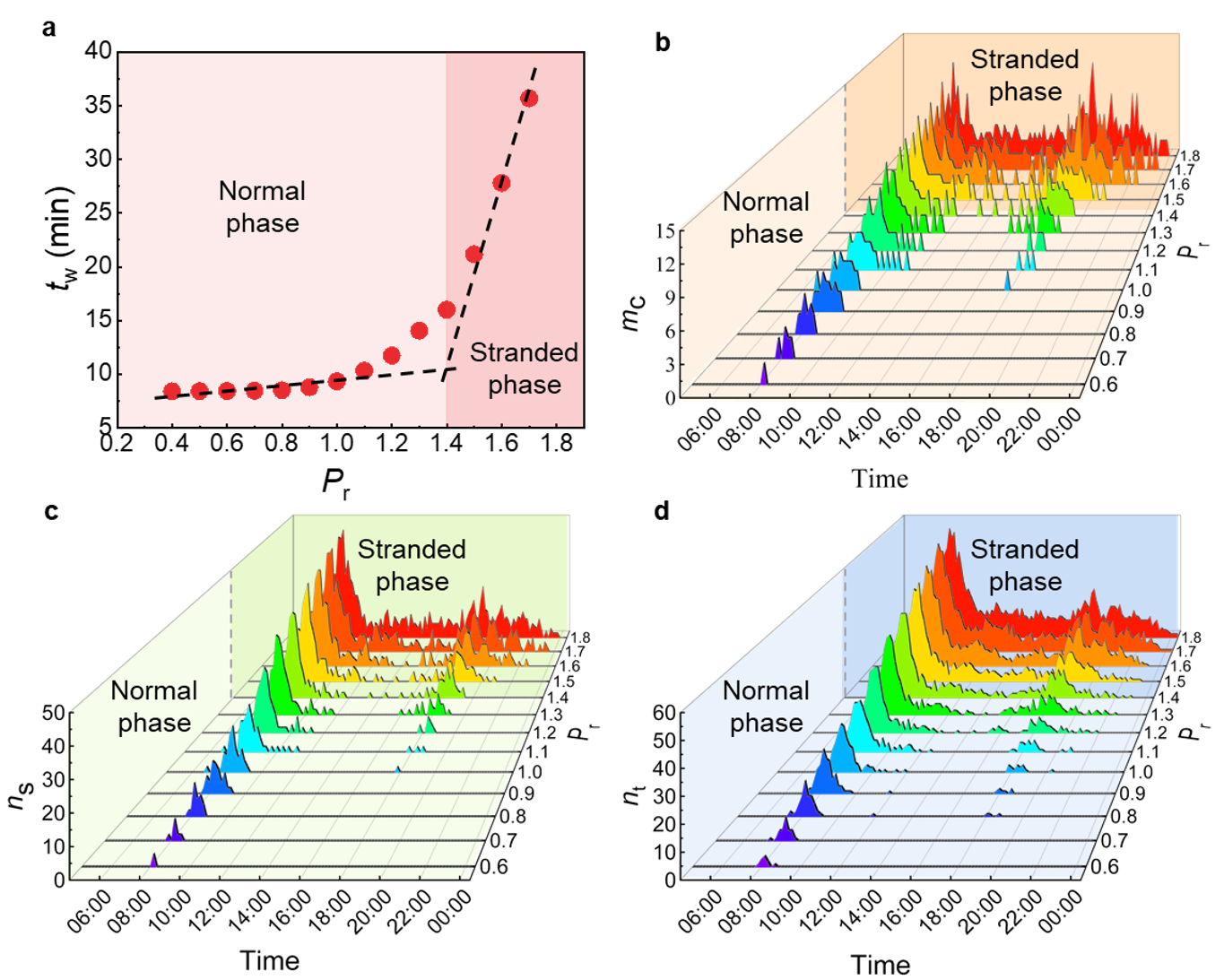}
	\caption{\textbf{a} Averaged passenger waiting time ($t_\mathrm{w}$) as a function of $P_\mathrm{r}$ which gives a transition point around $p_\mathrm{r} = 1.4$. \textbf{b}-\textbf{d} Temporal evolution of $m_\mathrm{c}$, $n_\mathrm{s}$, and $n_\mathrm{t}$ throughout the day as functions of $P_\mathrm{r}$ which show rush hours of morning and evening. The transition from Normal phase to Stranded phase can be determined by whether there is a correlation between the 2 peak periods.}
	\label{fig:fig4}
\end{figure}

In the context of spatiotemporal heterogeneity driven by real-system data, both passenger waiting time and 3 order parameters fail to reach steady states, exhibiting dynamic fluctuations over time. Consequently, the method to determining the transition point of the system, as depicted in Fig. \ref{fig:fig2}, becomes unfeasible. Alternatively, to characterize the non-equilibrium responsive behaviors of the system, one can introduce non-steady-state dynamics by increasing the demand level in a system that is in a steady state and measuring the relaxation time required for the system to establish a new steady state. In this study, the morning rush hour and evening rush hour (Supplementary Information, B.3.1, Fig.~B.7) are considered as 2 successive stimuli imposed on the system. Prior to the morning rush hour, the system is in a steady state where no stations with stranded passengers are observed. If the relaxation time of the system after the stimulus of morning rush hour is shorter than the time interval between the morning rush hour and evening rush hour, the system consistently returns to a steady state before the evening rush hour begins. This state is referred to as the ``normal phase" in present work. Conversely, if the relaxation time is shorter than the time interval between the morning rush hour and evening rush hour, the accumulation of stranding phenomena within the system leads to a continuous decline in service efficiency. This state is referred to as the stranded phase. By varying the passenger demand scale ($P_\mathrm{r}$), the changes in average passenger waiting time and the 3 order parameters over time are plotted to observe the system's responsive behaviors, as presented in Fig.~\ref{fig:fig4}. The average passenger waiting time directly reflects the system's service efficiency. To illustrate the phase transition characteristics, Fig.~\ref{fig:fig4}a presents the summation of the average passenger waiting time over the entire day as a function of $P_\mathrm{r}$. Importantly, a similar transition point, as shown in Fig.~\ref{fig:fig2}a, is observed at $P_\mathrm{r} \approx 1.4$, which can be defined as the threshold for resilience failure. When the demand scale is below this threshold, such as when $P_\mathrm{r} = 1.0$ representing the demand scale in a real normal day, 3 order parameters reflecting the level of stranded passengers exhibit distinct morning and evening rush hour characteristics with 2 peaks in the temporal dimension. Moreover, these 2 peaks do not overlap, creating a clear time window where no stranding phenomenon occurs. This finding implies that under normal demand scale for the morning rush hour and evening rush hour stimulus, the existing subway service effectively suppresses the occurrence of PPS within a limited time, without correlating to the subsequent rush hour. As the demand scale decreases, both the peak values of the 3 order parameters for the morning rush and evening hour diminish. When $P_\mathrm{r} < 1.0$, only the morning rush hour exhibits stations with stranded passengers and stranding clusters, while the evening rush hour is nearly absent. As $P_\mathrm{r}$ increases, the relaxation time for the morning rush hour extends, and the start time of the evening rush hour is advanced. After $P_\mathrm{r}$ reaches 1.4, the stranded station clusters generated by the morning rush hour fail to dissipate completely before the onset of the evening rush hour. Consequently, there is no time window without stranding between the morning rush hour and evening rush hour, making the subway network incapable of providing effective service due to PPS. This observation contrasts with the distinct dynamic characteristics of the morning rush and evening hour that are clearly separated when $P_\mathrm{r} < 1.4$. Hence, we define $P_\mathrm{r} = 1.4$ as the transition point of this system, serving as a metric for measuring the threshold of service resilience failure.

\subsection{Dynamics mechanism of cluster growth}
Understanding the dynamic mechanism behind PPS generation is crucial for understanding the phase transition like behaviors in service systems and holds significant practical implications for enhancing the resilience of the subway. The emergence of PPS in complex dynamical systems, comprising demand and supply basic units, is attributed to the complex and interconnected factors. In this study, the eigen microstate method of non-equilibrium statistical physics is used to uncover the characteristic dynamic patterns of PPS formation (see “Method”for details), which not only helps to understand the formation of clusters but also facilitates the study of cluster vanishing. include cluster formation and vanishing. According to the singular value decomposition (SVD)\cite{38}, the spatiotemporal correlation matrix $\bm{A}$ of stranded passenger number at each station can be factorized as $\bm{A} = \bm{U}\cdot\bm{\Sigma}\cdot \bm{V}^\mathrm{T}$, where $\bm{U}$ is the eigen microstate and $\bm{V}$ is their evolution with time, $\bm{\Sigma}$ is a diagonal matrix with the elements of $\sigma_I$. By equation ${{\bm{A}}_I^\mathrm{e}}\equiv \bm{U_I}\bigotimes \bm{V_I}$, we get ${{\bm{A}}_I^\mathrm{e}}$. We call the ensemble defined by ${{\bm{A}}_I^\mathrm{e}}$ an eigen ensemble of system, where $I$ represents the index of eigenstates. We consider $\sigma_I$ as the probability amplitude and $\omega_I = \sigma_I^2$ as the probability of the eigen ensemble ${{\bm{A}}_I^\mathrm{e}}$ in the statistical ensemble $\bm{A}$. The larger the probability $\omega_I$ associated with an eigen microstate, the greater its contribution to the formation of the PPS phenomenon. The eigen patterns with the largest eigenvalues are crucial factors determining the PPS phenomenon. By comparing the spatiotemporal characteristics of these key eigen microstates with the spatiotemporal characteristics of supply and demand, the micro mechanisms underlying the formation of these eigen patterns can be identified. Considering that the 4 commuting corridors are independent in the spatial dimension, the eigen microstate method analyzing is applied separately to analyze Clusters \uppercase\expandafter{\romannumeral1}-\uppercase\expandafter{\romannumeral6}.

\begin{figure}[!ht]
	\centering
	\includegraphics[width=0.85\textwidth]{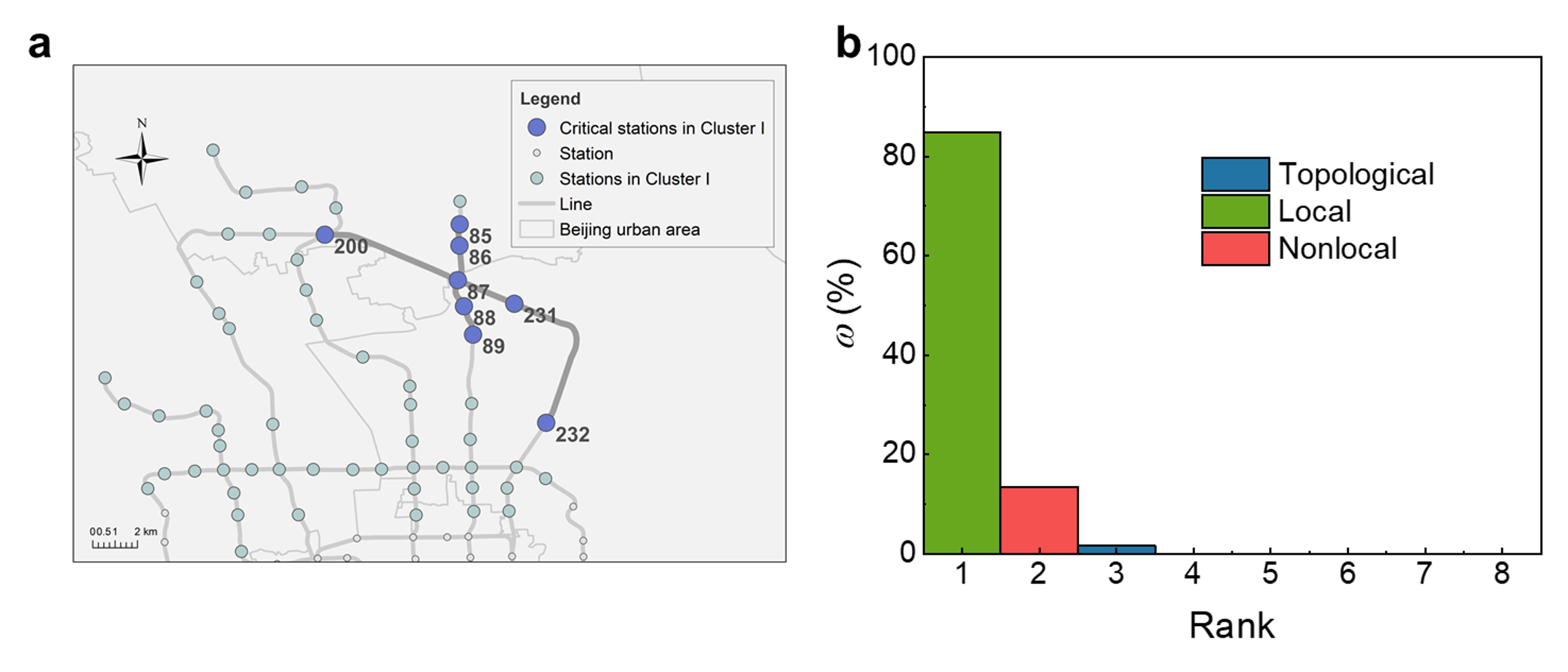}
	\caption{Station distribution and probabilities of eigen microstates for Cluster \uppercase\expandafter{\romannumeral1}. \textbf{a} Distribution of stations and critical stations within Cluster \uppercase\expandafter{\romannumeral1}. The labeled critical stations include Station 200 (Huoying), Station 87 (LisuiQiao), Station 231 (BeiYuan), Station 232 (WangjingXi), Station 86 (Tiantong Yuan), Station 85 (Tiantong Yuan), Station 88 (Lishuiqiao Nan), and Station 89 (BeiyuanluBei). \textbf{b} Probabilities of the first 8-th eigen microstates corresponding to Cluster \uppercase\expandafter{\romannumeral1} during 6:00-13:00.
	}
	\label{fig:fig5}
\end{figure}

Taking Cluster \uppercase\expandafter{\romannumeral1} (Fig.~\ref{fig:fig5}a) of the northern commuting corridor as an example, this corridor serves 2 adjacent large residential centers, Huilongguan and Tiantongyuan, collectively known as the "Huitian area" for commuting demands. This area covers approximately 63 square kilometers and accommodates nearly 900,000 residents, earning the title of "Asia's largest sleeping town". A large number of residents commute between the Huitian area and the city center by taking subway as their daily transportation. We concentrate on the morning rush hour, from 6:00 to 13:00 to reveal the formation and vanishing of Cluster I. We take a time snapshot every 10 minutes to record the number of stranded passengers at each station in Cluster I, resulting in a spatiotemporal correlation matrix of stranded passenger counts with 43 time snapshots. By performing SVD on this matrix, we obtain 8 independent eigen microstates ${{(A}_I^e)}_{it}$ and their eigen values $\omega_I$. 8 $\omega_I$  are shown in Fig.~\ref{fig:fig5}b  , and the sum of probabilities of the top three micro eigenstates reaches 99.983\%. We consider these 3 patterns, named as the Local Pattern, Nonlocal Pattern, and Topology Pattern based on their potential causes (more details in Supplementary Information, B.4.1), as the major mechanisms of the PPS phenomenon. They provide explanations for the impact of passenger demand on PPS, the propagation mechanism of PPS, and the impact of network topology on PPS. Similar analyses are conducted for Cluster \uppercase\expandafter{\romannumeral2}-\uppercase\expandafter{\romannumeral4}, as well, leading to evolution mechanisms similar to Cluster \uppercase\expandafter{\romannumeral1} (see Supplementary Information, B.4.2, Fig.~B.12 - Fig.~B.14). It is worth noting that the eigen microstate method can be regarded as an unsupervised machine learning method that excavates hidden patterns from the Spatiotemporal data. The interpretations of these patterns often require a phenomenological and empirical perspective\cite{39}. Therefore, while the eigen microstate method used in this study provides rigorous eigen modes, the interpretation of the underlying micro mechanisms and the naming of these patterns are somewhat subjective and serve as a reference.

% \vspace{-0.2cm}
% \vspace{-0.2cm}

\subsection{The impact of supply and demand on cluster growth}

\begin{figure}[!ht]
	\centering
	\includegraphics[width=0.7\textwidth]{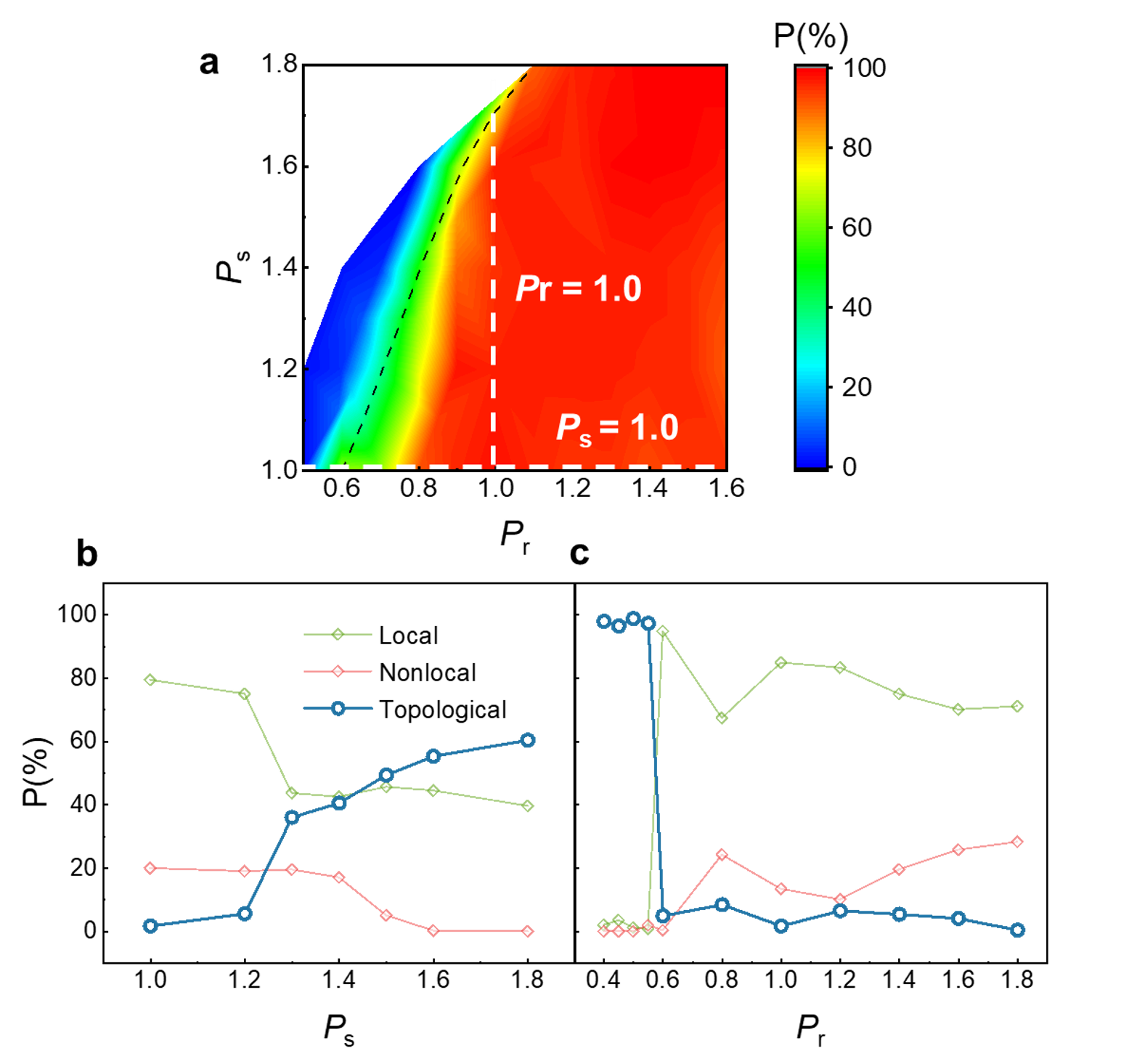}
	\caption{\textbf{a} Contribution of Topological pattern depending on scales of supply and demand. \textbf{b} Contributions of Local pattern, Nonlocal pattern and Topological pattern in the total variance for different values of $P_\mathrm{s}$. \textbf{c} Contributions of Local pattern, Nonlocal pattern and Topological pattern in the total variance for different values of $P_\mathrm{r}$.}
	\label{fig:fig10}
\end{figure}

The equilibrium between supply and demand is a fundamental factor that determines the state of the system. Therefore, we investigate the changes in the influence of different patterns by manipulating the demand scale and supply scale. In the present study, we maintain the spatiotemporal characteristics of the supply and the ratio of scheduled train departures to actual train departures is used as the system's service supply scale, denoted as $P_\mathrm{s}$. The details of the train departure frequency regulation scheme can be found in the Supplementary Information, B.4.3. The eigen microstate method is used to analyze the Local, Nonlocal, and Topological patterns of Cluster \uppercase\expandafter{\romannumeral1}. Fig.~\ref{fig:fig10}a illustrates the proportional contribution of Topological pattern depending on both $P_\mathrm{r}$ and $P_\mathrm{s}$ which equals to the combined Local and Nonlocal patterns by considering the normalization condition of these 3 contributions. The blue area represents supply surplus, while the red area represents supply shortage. In the region where $P_\mathrm{r}$ is relatively small and $P_\mathrm{s}$ is relatively large, the system does not exhibit stranding due to excessive supply. Thus, eigen microstate method is not applicable, as indicated by the blank area in Fig.~\ref{fig:fig10}a. The black dashed line in the green region corresponds to the boundary of the combined changes in the Local and Nonlocal patterns. Below this boundary, the Topological pattern dominans the PPS phenomena, and the influence of the Local and Nonlocal patterns is minimal. With an increasing ratio of $P_\mathrm{s}$ over $P_\mathrm{r}$, the influence of the Local and Nonlocal patterns gradually amplifies, confirming that these 2 patterns arise from the mismatch between supply scale and demand scale. 

\subsubsection{The impact of supply scale on patterns}

To observe the impact of supply scale regulation on stranding phenomenon, we consider Cluster \uppercase\expandafter{\romannumeral1} as an example. Under the condition with constant demand scale, we increase the supply scale on Line 5 and Line 13, where Cluster \uppercase\expandafter{\romannumeral1} is located, to enhance the supply scale while maintaining the spatiotemporal distribution characteristics of the supply. This allows us to observe the inhibitory effect of increased supply scale on the Local and Nonlocal patterns caused by demand surplus (see Fig.~\ref{fig:fig10}b), corresponding to the white dashed line parallel to the x-axis in Fig.~\ref{fig:fig10}a at $P_\mathrm{r} = 1.0$. As shown in Fig.~\ref{fig:fig10}b, with the increase in the supply scale, the contribution of the Topological pattern continues to rise, while both the Local and Nonlocal patterns are suppressed. In Fig.~\ref{fig:fig10}b, the increase in the contribution of the Topological pattern comes at the expense of decreased contributions from the Local and Nonlocal patterns. This indicates the validity of associating the Local and Nonlocal patterns with supply shortages.

When $P_\mathrm{s} > 1.5$, the Nonlocal pattern, which is an important mechanism in the PPS, is almost completely suppressed, and the contribution of the Local pattern is significantly reduced. Compared to the Nonlocal pattern, the Local pattern is more sensitive to changes in the supply scale. When $P_\mathrm{s} \approx 1.3$, the contribution of the Local pattern decreases rapidly, indicating that the increase in supply scale directly satisfies the supply. Meanwhile, the contribution of the Nonlocal pattern decreases slowly. When $P_\mathrm{s} \in [1.3, 1.5]$, the Local and Nonlocal patterns exhibit mutual influences. With the increase of supply scale, the Nonlocal pattern shows a weak peak that initially increases and then decreases, while the Local pattern exhibits a valley. It is important to note that with the increase in supply, the absolute of stranding scale decreases, but the contributions of the Local and Nonlocal patterns slightly increase. The Local pattern is highly sensitive to supply change, leading to a decrease as demand increases. The trigger for the Nonlocal pattern is the stranding generated by the Local pattern, allowing it to propagate downstream through the PPS mechanism. Therefore, the regulation of the Nonlocal pattern by supply scale is indirect, with the response to an increase in supply lagging behind that of the Local pattern. It is worth noting that when supply exceeds 1.5, the contribution of the Nonlocal pattern is almost completely eliminated, while the improvement of the Local pattern becomes slower with further increases in the supply scale.

The above results validate the rationality of the cluster growth mechanism analyzed earlier and provide valuable insights for regulatory policies. Increasing the supply within a certain range is advantageous for suppressing both the Local and Nonlocal patterns. The influence on the Nonlocal pattern gradually strengthens as the supply increases to the range of 1.3-1.5. However, once the supply exceeds 1.5, the inhibitory effect of the Local pattern significantly diminishes, and the Nonlocal pattern is completely eliminated. Increased transportation supply has limited impact on suppressing the Local and Nonlocal patterns. To further enhance the system's service efficiency, it is recommended to shift the focus of regulatory policies towards optimizing the network's topological structure.

\subsubsection{The impact of demand scale on patterns}

In the context of a constant supply, accommodating passenger demand is also an important issue. Therefore, we maintain the supply at its actual scale and regulate the demand scale $P_\mathrm{r}$, while observing the contributions of the 3 patterns. The results are presented in Fig.~\ref{fig:fig10}c which corresponds to the white dashed line parallel to the y-axis line in Fig.~\ref{fig:fig10}a representing $P_\mathrm{s} = 1.0$. The demand regulation method aligns with the approach described in section 2.3.1. As illustrated in Fig.~\ref{fig:fig10}c, after $P_\mathrm{r}$ exceeds 0.6, an increase in $P_\mathrm{r}$ leads to the Local and Nonlocal patterns becoming the dominant factors contributing to passenger stranding. Both the Local and Nonlocal patterns are enhanced, while the influence of the Topological pattern diminishes significantly, with its contribution nearly eliminated. This further confirms that the Local and Nonlocal patterns originate from the matching between demand and supply. The results suggest that the supply scale is effective in suppressing the Local and Nonlocal patterns when the demand is high, particularly in suppressing the Local pattern. When the demand scale is less than 0.6, increasing the supply has limited impact on suppressing the Local and Nonlocal patterns; therefore, alternative approaches, such as modifying the network structure to enhance system efficiency, should be considered.

Furthermore, we observed that as $P_\mathrm{r}$ increases, the Topological pattern and the Local pattern of Cluster \uppercase\expandafter{\romannumeral1} respectively dominate at different stages, while the influence of the Local pattern in Clusters \uppercase\expandafter{\romannumeral2}-\uppercase\expandafter{\romannumeral4} remains significant. This is due to the fact that Cluster \uppercase\expandafter{\romannumeral1} was constructed earlier and has a relatively complex topological structure. Stations with high demand and transfer stations are located in close proximity, resulting in the combined impact of multiple patterns on passenger stranding. Moreover, during the process of changing $P_\mathrm{r}$, the Topological pattern in this cluster has a relatively larger contribution compared to other clusters. On the other hand, the TOD pattern is more pronounced in Clusters \uppercase\expandafter{\romannumeral2}-\uppercase\expandafter{\romannumeral4}, which consist of a larger number of non-transfer stations. Consequently, the Local pattern caused by demand remains prominent in these clusters (for the analysis of the evolution mechanism in Clusters \uppercase\expandafter{\romannumeral2}-\uppercase\expandafter{\romannumeral4}, please refer to SI, 2.4.5, Fig. S13, Fig. S14).

\section{Discussion}
In this study, we examine the subway system as a complex system that consists of 2 types of basic units, namely passengers and subway trains with a service relationship. By employing computational simulations of a big-data-driven model system, we explore the dynamics of stranding phenomenon within the subway system, particularly how it evolves in response to variations in the scale of supply. Our computational results unveil analogous behaviors of phase transitions in both service efficiency, as quantified by the average waiting time for passengers, and the scale of stranding phenomenon within the subway. We identify a transition point, which quantitatively characterizes the threshold of subway service resilience failure. Employing the method of eigen microstates from non-equilibrium statistical physics, we elucidate the formation mechanisms of the PPS phenomenon. These achievements enhance the understanding of the features of service efficiency in subway networks and provides a theoretical foundation for strategies such as topological planning and optimization of train schedules to improve service. The proposed framework, involving complex systems comprising multiple interdependent basic units with service relationships, offers a valuable perspective for studying urban systems. In the pursuit of smart cities, this framework serves as a reference for evaluating urban planning, governmental policies, and business decisions. Furthermore, the methodology used in the present work has broader implications for similar scheduling and logistics management challenges encountered in other service systems, including demand-driven or flexible bus scheduling, elevator dispatching in skyscrapers, supply chain management, medical appointment assignment, logistics management for aircraft carriers, and container logistics management.

\section{Method}
We begin by adopting Gibbs' ensemble method \cite{20} and define a statistical ensemble comprising $M$ microstates from a complex system with $N$ components. This statistical system is represented by a normalized $N\times M$ ensemble matrix $\bm{A}$, in which each row corresponds to the number of stranded passengers $S_i$($t$) ($i$ = 1, $M$) at different stations for a specific time snapshot. In this context, $M$ is set to 234, while the number of time samples is $N$ = 43.

At time $t$, the number of stranded passengers at station i is denoted as SPD, which represents the quantity of passengers. To calculate the average value of SPD at station $i$ over $M$ time periods, we compute the mean value as
\begin{equation}
	\langle S_i \rangle = \frac{1}{M}\sum_{i=1}^{M}S_i(t).
\end{equation}

At a certain time $t$, the agent $i$ has a fluctuation
\begin{equation}
	\delta S_i(t) = S_i(t) - \langle S_i \rangle.
\end{equation}
We define a microstate with fluctuations of all agents, which is represented by a normalized $N$-dimensional vector\cite{21}
\begin{equation}
	\delta \bm{S}(t) = \begin{bmatrix}
		\delta S_1(t) \\
		\delta S_2(t) \\
		\vdots \\
		\delta S_N(t)
	\end{bmatrix}.
\end{equation}

With the $M$ microstates, we can compose a statistical ensemble of the complex system. This ensemble is described by an $N \times M$ matrix $\bm{A}$ with elements
\begin{equation}
	A_{it} = \frac{\delta S_i(t)}{\sqrt{c_0}},
\end{equation}
where $C_0 = \sum_{t=1}^{M}\sum_{i=1}^{N}\delta S_i^2(t)$. The column order of $\bm{A}$ is in accord with the evolution of microstate.
As in Reference\cite{40}, the correlation between the microstates at $t$ and $t^{\prime}$ is defined by their vector product
\begin{equation}
	C_{tt^\prime} = \delta S\left(t\right)^\mathrm{T}\cdot\delta S\left(t^\prime\right) = \sum_{i=1}^{N}{\delta S_i\left(t\right)\delta S_i\left(t^\prime\right)}.
\end{equation}

With $C_{tt^\prime}$ as its elements, we can get an M × M correlation matrix of microstate as
\begin{equation}
	\bm{C} = C_0\bm{A}^\mathrm{T}\cdot \bm{A},
\end{equation}
whose trace $\mathrm{Tr}C = \sum_{t=1}^{M}C_{tt} = C_0$.

According to the singular value decomposition (SVD) \cite{41}, the ensemble matrix $\bm{A}$ can be factorized as
\begin{equation}
	\bm{A} = \bm{U}\cdot\bm{\Sigma}\cdot \bm{V}^\mathrm{T},
\end{equation}
where $\delta$ is an $N \times M$ diagonal matrix with elements
\begin{equation}
	\delta_{ij} = \begin{cases}
		\delta_I, & \text{for $i = j \le r$} \\
		0, & \text{for otherwise},
	\end{cases},
\end{equation}
where $r = min\left(N,M\right)$.

We can rewrite the ensemble matrix $\bm{A}$ as
\begin{equation}
	\bm{A} = \sum_{I=1}^{r}{\sigma_I \bm{A_I}^\mathrm{e}},
\end{equation}
where $\bm{A_I}^\mathrm{e}\equiv \bm{U_I}\bigotimes \bm{V_I}$ is an $N$ $\times$ $M$ matrix with elements ${{(A}_I^\mathrm{e})}_{it} = U_{it}V_{it}$. We call the ensemble defined by $A_I^e$ an eigen ensemble of system.

From ${\mathrm{Tr}}C = C_0$, we have the relation
\begin{equation}
	\sum_{I=1}^{r}\sigma_I^2 = 1.
\end{equation}

Therefore, we consider $\sigma_I$ as the probability amplitude and ${\omega_I} = \sigma_I^2$ as the probability of the eigen ensemble $\bm{A_I}^\mathrm{e}$ in the statistical ensemble $\bm{A}$. This is analogous to quantum mechanics, where a wave function can be written in terms of eigen functions. The square of the absolute value of expansion coeﬀicient is the probability of the corresponding eigen microstate.

	\section*{Acknowledgements}
	This work was supported by the Beijing Social Science Foundation (No. 21JCB093), Major Program of The Humanity and Social Science Fund of Beijing Jiaotong University (No.2023JBW1002) and National Key Research and Development Program of China (NKPs; Nos. 2021YFA1000300 and 2021YFA1000303).
	
	\section*{Author contributions}
	Xinghua Zhang conceived the idea. Xinghua Zhang and Xinyi Li developed the methodology. Shengda Zhao implemented the computer code and supported the algorithms. Liang Wang applied statistical and mathematical techniques to analyze the study data. Qing Wang prepared the published work, specifically focusing on visualization. Liang Wang and Daqing Gong provided the data. Xun Zhang,Fang Liu and Xiaodong Zhang created the models. Xinghua Zhang and Xinyi Li wrote the paper. All authors have reviewed and approved the manuscript.

	\section*{Competing interests}
The authors declare no competing interests.

% % \section{appendix_main}
% \documentclass{article}
% \usepackage[fontsize=12pt]{fontsize}
% \usepackage{hyperref}       % hyperlinks
% \hypersetup{colorlinks = true,  
% 	linkcolor = blue, 
% 	urlcolor = blue,%网页链接为蓝色
% 	citecolor = blue} %文献引用颜色设置为蓝色
% \usepackage[verbose=true,letterpaper]{geometry}

% \AtBeginDocument{
% 	\newgeometry{
% 		textheight=9in,
% 		textwidth=7in,
% 		top=1in,
% 		headheight=14pt,
% 		headsep=25pt,
% 		footskip=30pt
% 	}
% }            
% \usepackage{setspace}
% \usepackage{multirow}
% \usepackage{makecell}
% \usepackage{graphicx}
% \usepackage[numbers,square,super,comma,compress]{natbib}
% \usepackage{booktabs}       % professional-quality tables
% \usepackage{amsfonts}       % blackboard math symbols
% \usepackage{amsmath,tabularx}
% \usepackage{float}

% \renewcommand{\thefigure}{\thesection.\arabic{figure}}  
% \renewcommand{\theequation}{\thesection.\arabic{equation}} 
% \renewcommand{\thetable}{\thesection.\arabic{table}} 

% \title{{\LARGE \textbf{Supplementary Information}}\\
% 	\textbf{Phase transition like behaviors of propagation of passenger stranding phenomena in subway networks}}

% \author{}
% \date{ }

% \begin{document}
% 	\maketitle 	
% 	\setstretch{2}
%  \thispagestyle{empty} 
%  \newpage
% \thispagestyle{empty} 

% \tableofcontents
% \thispagestyle{empty}   %目录页无页码

% \newpage
% \setcounter{page}{1}  

\appendix
\section{Data and system}
\subsection{Data description and processing}\label{sec:A.1}
\subsubsection{Passenger-related data}
 \textbf{(1) Data description}
 
 In this study, the passenger demand dataset for the system II is obtained from the Automatic Fare Collection (AFC) system of Beijing Rail Transit. The dataset comprises passenger travel data collected by the AFC system and includes all travel information from May 10th to May 14th, 2021. The data attributes include TXN DATE TIME, DEVICE LOCATION, CARD SERIAL NUMBER, TRIP ORIGIN LOCATION, ENTRY TIME, PRODUCT ISSUER ID, PRODUCT TYPE, PAYMENT VALUE, CARD LIFE CYCLE COUNT, RECONCILIATION DATE, SETTLEMENT DATE, SAM ID, DEVICE ID, SOURCE PARTICIPANT ID, PURSE REMAINING VALUE, PURSE\_REMAINING\_VALUE, PTSN. The attributes of the data used in the study include TXN DATE TIME, DEVICE LOCATION, CARD SERIAL NUMBER, TRIP ORIGIN LOCATION, ENTRY TIME.
 
 \textbf{(2) Data processing}
 
In this study, the network model comprises 9 out of the total 25 lines of the Beijing subway network as of December 2021. To process the original dataset, the following steps were taken:

(a) Remove records with missing TRIP ORIGIN LOCATION, ENTRY TIME, TXN DATE TIME and DEVICE LOCATION.

(b) Remove records which TRIP ORIGIN LOCATION and DEVICE LOCATION are the same.

(c) When multiple records have the same PRODUCT ISSUER ID, TRIP ORIGIN LOCATION, ENTRY TIME, TXN DATE TIME and DEVICE LOCATION as other records, keep only one of them.

(d) Remove records where the ENTRY TIME or TXN DATE TIME falls outside the operating hours of the subway (from 4:30 a.m. to 0:30 a.m. the next day).

(e) Remain records where both TRIP ORIGIN LOCATION and DEVICE LOCATION are limited to stations along Lines 1, 2, 4, 5, 6, 7, 8, 10, and 13.

 A total of 15,850,040 passenger records were remained, including 3,811,418 records for Monday, 3,831,076 records for Tuesday, 3,833,882 records for Wednesday, 3,799,731 records for Thursday, and 4,023,933 records for Friday.

\subsubsection{Subway-related data}
 \textbf{(1) Data description}

(a) Categorization of subway trains

Categorized by type, the trains in Beijing are classified into categories A, B, C, D, and L. In our system, all trains belong to type B, with a maximum capacity of 250 passengers per train. In system I, the passenger capacity of each train is set at 1500 passengers. In system \uppercase\expandafter{\romannumeral2}, the trains on each line are identical, but there may be slight variations in the maximum passenger capacity among different lines due to the train type and formation. The train types, compositions, and maximum passenger capacities for each line are listed in Table \ref{tab:table1}. The actual train capacity in system \uppercase\expandafter{\romannumeral2} needs to be adjusted through parameter tuning, and the tuning results will be presented in the following sections.

\begin{table}[!ht]
	\caption{Train Type, Formation, and Maximum Passenger Capacity Information}
	\centering
	\begin{tabular}{ccc}
		\toprule
		Line     & Formation     & Maximum passenger capacity \\
		\midrule
		1 & 6  & 1500     \\
		2 & 6  & 1500     \\
		4 & 6  & 1500     \\
		5 & 6  & 1500     \\            
		6 & 8  & 2015     \\		
		7 & 8  & 2015     \\
		8 & 6  & 1500     \\
		10 & 6  & 1500     \\
		13 & 6  & 1500     \\
		\bottomrule
	\end{tabular}
	\label{tab:table1}
\end{table}

Categorized by shuttle interval, the trains in the system can be classified as shuttle trains and non-shuttle trains. System \uppercase\expandafter{\romannumeral1} exclusively consists of shuttle trains, while the composition of trains in System \uppercase\expandafter{\romannumeral2} is determined based on real data.

(b) Train timetable data

The train timetable in System \uppercase\expandafter{\romannumeral2} is set according to actual timetable. The number of train departures on each line are shown in Table \ref{tab:table2}. 
The train timetable data utilized in our study was obtained through web scraping of real-time data from the Beijing Rail Transit App. Beijing Rail Transit App is maintained and released by the Beijing subway operating authority, provides comprehensive subway timetable information. The timetables include essential details such as train departure and arrival times, station specifics, and train departure frequencies for each line in system \uppercase\expandafter{\romannumeral2}.

\begin{table}[!ht]
	\caption{The total number of train departures on each line}
	\centering
		\begin{tabularx}{0.9\textwidth}{p{2cm} p{1.1cm} p{1.1cm} p{1.1cm} p{1.1cm} p{1.1cm} p{1.1cm} p{1.1cm} p{1.1cm} p{1.1cm}}
			\toprule
			Line & 1 & 2 & 4 & 5 & 6 & 7 & 8 & 10 & 13 \\
			\midrule
			Number & 652 & 543 & 667 & 594 & 614 & 760 & 394 & 495 & 574\\
			\bottomrule
		\end{tabularx}
	\label{tab:table2}
\end{table}

 \textbf{(2) Data processing}
 
Based on the acquired train timetable data, the departure time for each train was extracted. Using the timing information for the arrival at 2 adjacent stations, the travel time between each pair of stations was calculated as the time interval. These intervals represent the travel time between adjacent stations.

\subsection{2 systems}\label{sec:A.2}

We have established 2 systems: an ideal system and a realistic system, each comprising 3 components: train operation, passenger generation, and train service to passengers.

\subsubsection{Train operation}

 \textbf{(1) Departure}

Once reached the departure time, both 2 systems automatically generate new trains from each originating station. The departure time, originating station, and destination station are set as follows:

(a) In	system I

At the beginning of the simulation, trains on each line start departing from their respective originating stations and continue until the end of the simulation, with a departure interval set at 4 minutes. The originating and destination stations for the trains are obtained from real data (Table \ref{tab:table3}), and each line has 2 travel directions. For linear lines, the originating and destination stations are located at the 2 ends of each line. For circular lines, the originating and destination stations are adjacent stations. The departure direction is from the originating station towards the destination station.

 \begin{table}[!ht]
 	\caption{Information of Line Types, Direction, Originating Stations, and Destination Stations}
 	\centering
 	\begin{tabular}{ccccc}
 		\toprule
 		Line & Type & Direction Type & Originating Stations & Destination Stations \\
 		\midrule
 		\multirow{2}{*}{1} & \multirow{2}{*}{Straight} & 1 & Gucheng & Universal Report \\
 		& & 2 & Universal Report & Gucheng \\
 		\multirow{2}{*}{2} & \multirow{2}{*}{Loop} & 1 & Jishuitan & Xizhimen \\
 		& & 2 & Xizhimen & Jishuitu \\
 		\multirow{2}{*}{4} & \multirow{2}{*}{Straight} & 1 & Anheqiaobei & Tiangongyuan \\
 		& & 2 & Tiangongyuan & Anheqiaobei \\
 		\multirow{2}{*}{5} & \multirow{2}{*}{Straight} & 1 & Tiantongyuanbei & Songjiazhuang \\
 		& & 2 & Songjiazhuang & Tiangongyuanbei \\
 		\multirow{2}{*}{6} & \multirow{2}{*}{Straight} & 1 & Tiantongyuanbei & Songjiazhuang \\
 		& & 2 & Songjiazhuang & Tiangongyuanbei \\
 		\multirow{2}{*}{7} & \multirow{2}{*}{Straight} & 1 & Liuliqiaodong & Universal Resort \\
 		& & 2 & Universal Resort & Liuliqiaodong \\
 		\multirow{2}{*}{8} & \multirow{2}{*}{Straight} & 1 & Yinghai & Zhuxingzhuang \\
 		& & 2 & Zhuxingzhuang & Yinghai \\
 		\multirow{2}{*}{10} & \multirow{2}{*}{Loop} & 1 & Bagou & Huoqiying \\
 		& & 2 & Huoqiying & Bagou \\
 		\multirow{2}{*}{13} & \multirow{2}{*}{Straight} & 1 & Xizhimen & Dongzhimen \\
 		& & 2 & Dongzhimen & Xizhimen \\
 		\bottomrule
 	\end{tabular}
 	\label{tab:table3}
 \end{table}
 
 (b) In system \uppercase\expandafter{\romannumeral2}
 
 The departure time, originating stations, and destination stations for both shuttle and non-shuttle trains on each line are obtained from real data. 
 
 \textbf{ (2) Travel process}

 (a) System \uppercase\expandafter{\romannumeral1}
 
\begin{figure}[!htbp]
 	\centering
 	\includegraphics[width=0.75\textwidth]{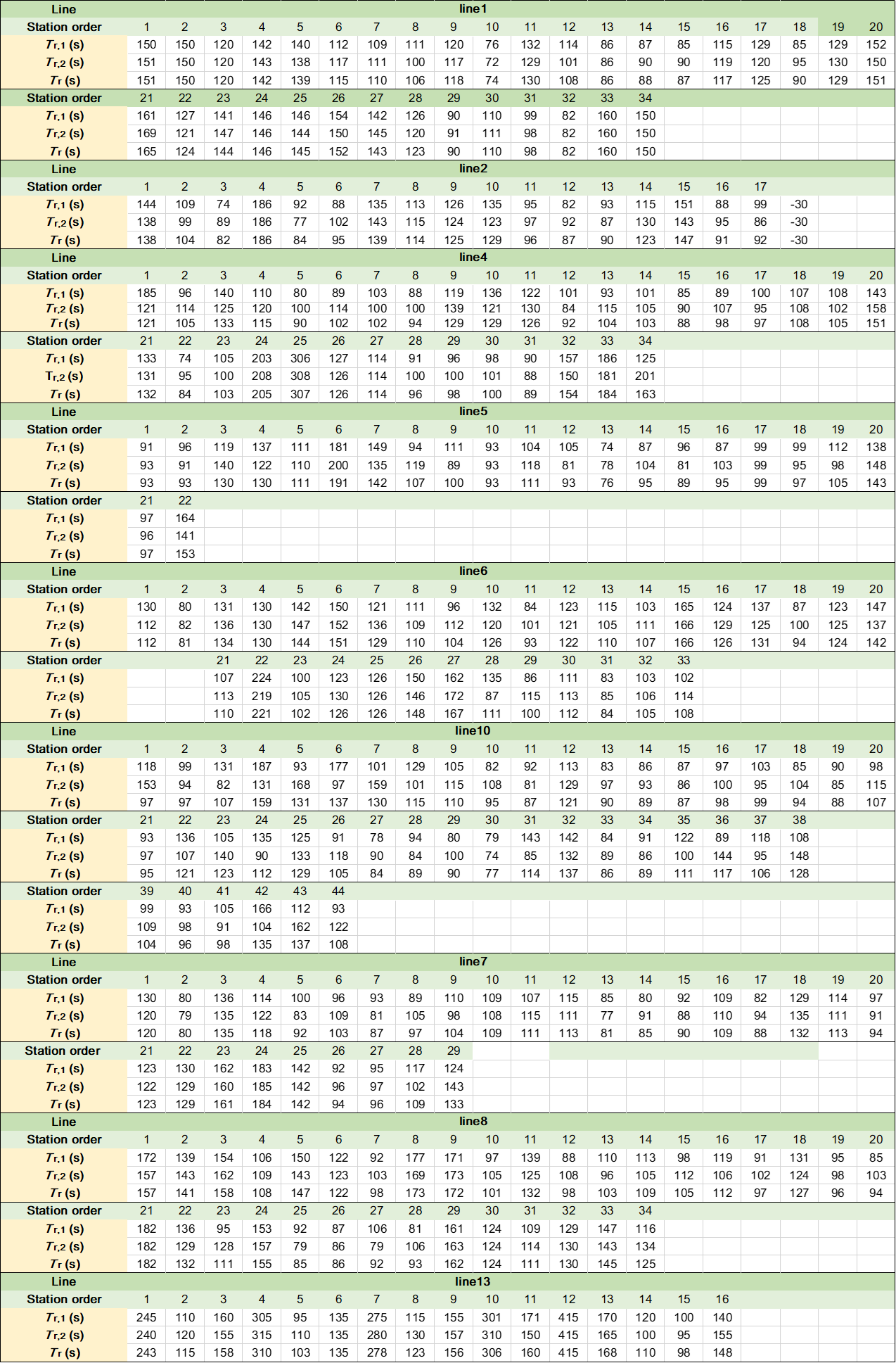}
 	\caption{Travel time between stations. The direction of lines in the table is arranged according to the direction from originating station to destination station (direction 1) on each line.}
 	\label{fig:figS1}
 \end{figure}

 The trains consistently travel on a single line, departing from the originating station and sequentially passing through each station between the originating station and the destination station. 
 Trains maintain a constant speed while traveling between adjacent stations. Since the distance between adjacent stations remains constant, the interval travel time can be determined using real data. The line segments between stations are represented as edges between nodes, and following the directions specified in Fig. \ref{fig:figS1}, each edge on each line is sequentially numbered as ``1, 2, 3, $\dots$" until all edges on the line are numbered. The average travel time taken by the trains to pass through is used as the interval travel time $T_i$ for the corresponding edges. Only intervals satisfying $T_\mathrm{i} \in (0, 1000]$ seconds are retained. The interval travel time consists of 2 components: the travel time of trains ($T_i$,0) and the time spent at stations ($T_\mathrm{s}$) (set as 30s). By subtracting the dwell time ($T_\mathrm{s}$) from the total interval, the travel time ($T_{i\mathrm{,1}}$ and $T_\mathrm{r,2}$) of trains on each edge of the line in both directions is obtained. The average travel times on corresponding edges in both directions are calculated to determine the travel time ($T_i$) of trains on each edge in the simulation system (Fig. \ref{fig:figS1}). This process starts from the departure and continues until all stations on the line would be traversed. Once the trains arrive at their destinations, they are removed from the system, and post-travel scheduling is not considered.
 
 \subsubsection{The generation of passengers}

 \textbf{ (1) Setting of passengers' entry time, originating stations, exit time and destination stations}
 
 The attributes of the data inputted into the system includes: ``TRIP ORIGIN LOCATION" ,``ENTRY TIME",`` TXN DATE TIME" and ``DEVICE LOCATION", which respectively represent passengers' originating stations, entry time, destination stations, and exit time. The following description outlines the simulation of passengers in both System \uppercase\expandafter{\romannumeral1} and System \uppercase\expandafter{\romannumeral2}.
 
 (a) In system I
 
 At the beginning of the simulation, System \uppercase\expandafter{\romannumeral1} generates $n_\mathrm{p}$ passengers every minute. The originating and destination stations for these passengers are randomly and independently selected from the entire pool of $N$ stations ($N = 234$). The passenger demand in the system is denoted as $P_\mathrm{u}$ and calculated as $n_\mathrm{p}/N$. 

 (b) In system \uppercase\expandafter{\romannumeral2}
  
 We randomly selected $N_\mathrm{d}$ records from the entire set of AFC data records, which comprised $N_\mathrm{R}$ records spanning from May 10th to May 14th, 2021. The selection was based on the records' attributes of entry time and passenger origin-destination (OD) pairs. These $N_\mathrm{d}$ selected records were used to generate passengers in System \uppercase\expandafter{\romannumeral2}. Given that $N_\mathrm{R}$ represents the total records from 5 working days ($N_\mathrm{R}=19,300,040$), we defined the number of passengers per day as $N\mathrm{r}=N_\mathrm{R}/5$. The demand scale id defined as $P_\mathrm{r}=N_\mathrm{d}/N_\mathrm{r}$. When the number of selected passenger records, Nd, equaled $N_\mathrm{r}$, the demand scale of passengers, $P_\mathrm{r}$, would be 1.0. When the simulation time reaches the entry time of a passenger, the system generates the corresponding passenger at the originating station, and the passenger immediately transitions to the waiting state.
 
   \textbf{(2) Passenger route planning}

 The Breadth-First Search (BFS) algorithm is employed to identify the $k$ ($k=5$) shortest paths for each passenger's OD pair. These paths are evaluated based on 3 factors: the number of stations passed ($n_\mathrm{s}$), the total travel time ($T = T_\mathrm{d}+ T_\mathrm{l}$, where $T_\mathrm{d}$ represents the total travel time and $T_\mathrm{l}$ represents the total station dwell time), and the number of transfer stations ($n_\mathrm{t}$). A scoring mechanism is applied to the $k$ paths, and the highest-scoring path is selected as the route for that OD pair. The scoring rule is defined as follows:
 
 \begin{equation}
 	Score = (10/n_\mathrm{s} + 1000/T + 1/n_\mathrm{t})/111.
 	\label{eq:eqS1}
 \end{equation}
 
 \subsubsection{Trains' service to passengers}\label{sec:A.2.3}
 
 \textbf{(1) Service process and rules}
 
 Before the simulation begins, the number of trains, timetables for each train, and maximum passenger capacity per train are predetermined and remain consistent throughout the simulation. When a train arrives, the following rules are applied in a specific order: passengers disembark before boarding, priority is given to passengers based on their arrival time, and if a train has reached its maximum capacity, no additional passengers can board. Specifically, for trains stopped at stations:
 
 (a) Passenger disembarkation
 
 If the train reaches its destination, all passengers onboard should disembark. Among the disembarking passengers, those whose disembarkation station differs from their intended destination station need to transfer, while others are simply removed. If the train has not reached its destination, passengers who have reached their intended destination station need to get off and are removed. Among the remaining passengers who have not reached their destination, those whose next station on their path differs from the next station the train will arrive at need to transfer. The transfer rule is as follows: after disembarkation, passengers who need to transfer must queue for their next train on the platform that aligns with both the passenger and the train's next station.
 
 (b) Passenger boarding
 
 If a passenger's next station matches the train's next station, they board the train in a first-come, first-served order. Otherwise, the passenger continues to wait in the queue.
 
 (c) Simulation duration
 
 System \uppercase\expandafter{\romannumeral1} duration is from 0 to 50400s and is collected every 5min.
 
 System \uppercase\expandafter{\romannumeral2} covered the period from 4:30 AM to 12:30 AM (a total of 73,800 seconds), with data collected every 10 minutes.
 
 \textbf{(2) Dynamic process of passenger boarding and disembarkation}
 
\begin{figure}[H]
 	\centering
 	\includegraphics[width=0.8\textwidth]{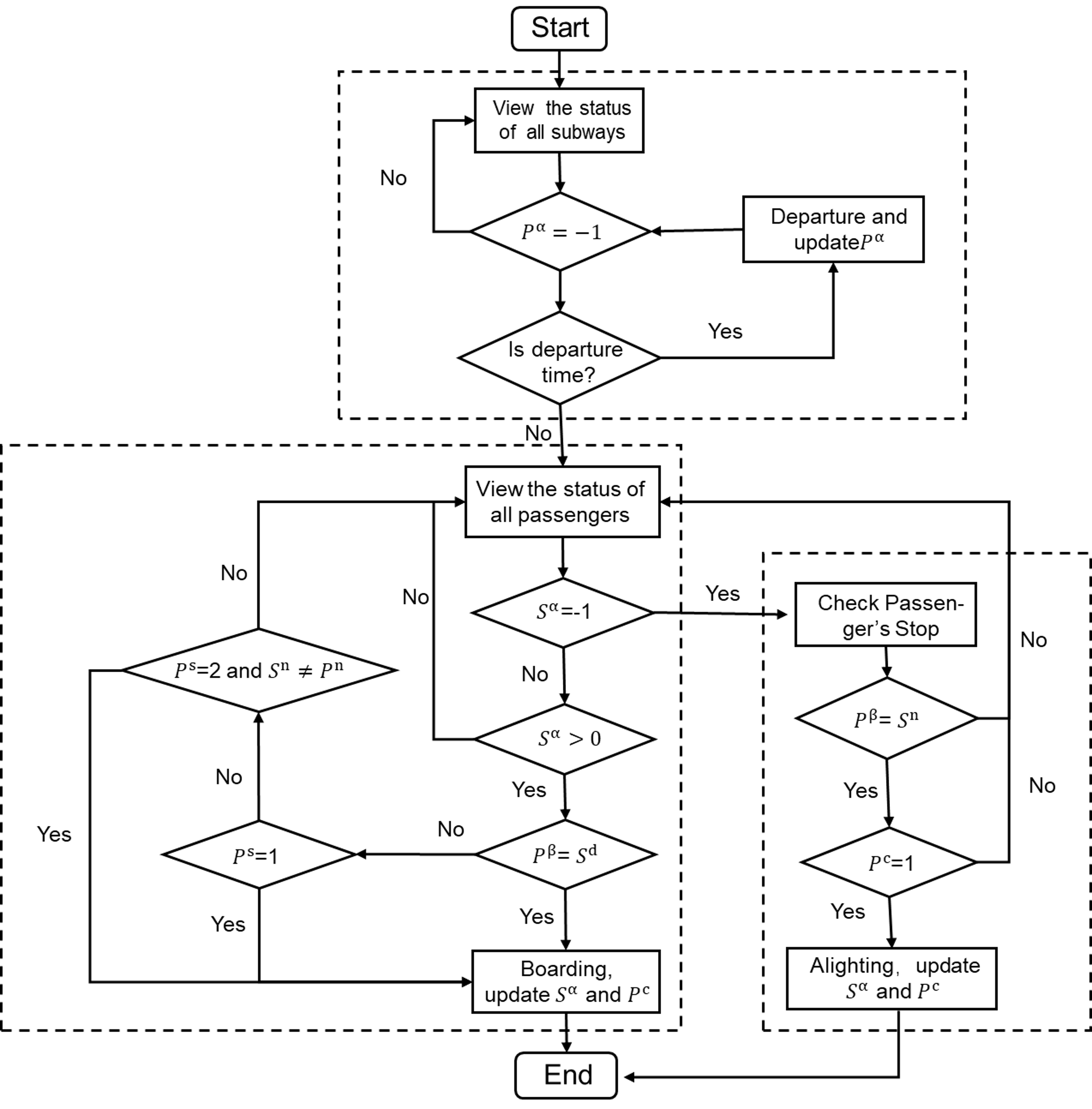}
 	\caption{Passenger boarding and disembarkation dynamics.}
 	\label{fig:figS2}
 \end{figure}

 The dynamic process of passenger boarding and disembarkation in the system can be illustrated by Fig. \ref{fig:figS2}, with specific symbols explained in Table \ref{tab:table4}.
 
We introduce the variable $P^\mathrm{\alpha}$ to denote the state of the train, while $P_\mathrm{b}$ represents the station where the train is currently positioned. A value of $P^\mathrm{\alpha} = 0$ indicates that the train has not yet departed, $P^\mathrm{\alpha} = -1$ signifies that the train is currently halted at a station, $P^\mathrm{\alpha} = 1$ denotes that the train is in transit, and $P^\mathrm{\alpha} = 2$ indicates that the train has reached its destination.
 
Passengers in the 2 systems are generated according to the following procedure. A passenger is generated every minute. Prior to the simulation, the passenger's entry time, originating station, and destination station are predetermined. During the simulation, the system employs the K-shortest path algorithm to determine the optimal travel route for each passenger based on their starting and ending stations. The 2 systems track the travel demand status ($S^\mathrm{\alpha}$) of all generated passengers at each moment. There are 4 states: $S^\mathrm{\alpha} = -1$ represents passengers waiting to board at the platform, $S^\mathrm{\alpha} > 0$ indicates that passengers have boarded and $S^\mathrm{\alpha}$ represents the train number they are on, and $S^\mathrm{\alpha} = 0$ denotes passengers who have reached their destination station.
 
 \begin{table}[!ht]
 	\caption{Notations, meaning and comments of the system}
 	\centering
 	\begin{tabular}{ccc}
 		\toprule
 		Notation & Meaning & Comments \\
 		\midrule
 		 {$P^\mathrm{\alpha}$} & {Trains' state} &  {\makecell[c]{$P^\mathrm{\alpha} = 0$: not yet departed; $P^\mathrm{\alpha} = -1$: \\ halted  at the station; $P^\mathrm{\alpha} = 1$: in transit; \\ $P^\mathrm{\alpha} = 2$: reached destination}} \\
 		
 	 {$S^\mathrm{\alpha}$} &  {Passengers' state} &  {\makecell[c]{$S^\mathrm{\alpha} = -1$: waiting to board; $S^\mathrm{\alpha} > 0$: \\  being onboard, and $S^\mathrm{\alpha}$ train number they are on; \\$S^\mathrm{\alpha} = 0$: reached destination station}} \\
 		 {$P^\mathrm{s}$} &  {Stations' type} &  {\makecell[c]{$P^\mathrm{s} = 1$: reached destination; $P^\mathrm{s} = 2$: \\reached at transfer station; $P^\mathrm{s} = 0$: otherwise}}\\
 		$P^\mathrm{\beta}$ & TrainS' location & Current station number of the train \\
 		$P^\mathrm{n}$ & Next Station of the Train & Next station number of the train\\
 		$P^\mathrm{c}$ & Train occupancy state & $P^\mathrm{c} = 1$: fully loaded; $P^\mathrm{c} = 0$: not fully loaded \\
 		$S^\mathrm{d}$ & Passengers' destination & Passengers’ destination\\
 		$S^\mathrm{n}$ & Next station of passengers & Next station of passengers\\
 		\bottomrule
 	\end{tabular}
 	\label{tab:table4}
 \end{table}
 
 \subsubsection{ Parameter settings and validation }\label{sec:A.2.4}
 It is important to note that not all lines are included in the network model in this study. Consequently, there may be passengers within the real traffic flow whose origins or destinations are not situated along the 9 lines of the network model. This discrepancy gives rise to deviations between the train capacity simulated by the system and the actual train capacity. To tackle this challenge, we leverage the entry and exit timestamps obtained from the AFC data to determine the passengers' travel time. By comparing the simulated travel time with the real travel time derived from the AFC data, we can make adjustments to the train capacity parameter, ensuring a more accurate representation of the system's capacity.
 
 As different types of trains on various lines, variations in train capacities arise across the network. The specific train capacities for each line are documented in Table \ref{tab:table1}. Throughout the parameter adjustment process, we ensure that the proportional relationship between train capacities on different lines remains consistent with the real train capacities observed for each line. In Table \ref{tab:table5}, we denote $q_0$ as the reference capacity of trains for Line 1, while $q$ represents the adjusted capacity of train cars for Line 1.
 
 \begin{table}[!ht]
 	\caption{\centering{Information on Reference Capacity and Design Standard Passenger Capacity of Subway Trains on Different Lines}}
 	\centering
 	\begin{tabular}{ccc}
 		\toprule
 		Line & Reference Capacity & Standard Capacity \\
 		\midrule
 		1 & 1500 & 1460 \\
 		2 & 1500 & 1460 \\
 		4 & 1500 & 1460 \\
 		5 & 1500 & 1460 \\
 		6 & 1500 & 1960 \\
 		7 & 2014 & 1960 \\
 		8 & 2014 & 1460 \\
 		10 & 2014 & 1460 \\
 		13 & 2014 & 1460 \\
 		\bottomrule
 	\end{tabular}
 	\label{tab:table5}
 \end{table}
 
 We calculated the fitting effect of different train capacity models by following steps.
 
 Step 1: We obtained the real boarding time $t_\mathrm{s,r}$ and disembarkation time $t_\mathrm{e,r}$ for each passenger from the AFC data. We also obtained the corresponding passenger boarding time $t_\mathrm{s,s}$($t_\mathrm{s,r} = t_\mathrm{s,s}$) and disembarkation time $t_\mathrm{e,s}$ from the system \uppercase\expandafter{\romannumeral2} (set $P_\mathrm{r} = 1.0$).
 
 Step 2: The real travel time for each passenger is given by $\Delta t_\mathrm{r} = t_\mathrm{e,r} - t_\mathrm{s,r}$, and the simulated travel time is given by $\Delta t_\mathrm{s} = t_\mathrm{e,s} - t_\mathrm{s,s}$. The absolute error between the simulated and actual travel times is calculated as $\Delta t = \Delta t_r - \Delta t_\mathrm{s}$.
 
 Step 3: We applied the 3-sigma criterion to remove outliers from the set of absolute errors.
 
 Step 4: The absolute time errors are normalized to obtain the relative error for each passenger, defined as $\Delta t' = \Delta t/\Delta t_\mathrm{r}$. A smaller value of $\Delta t'$ indicates that the simulation results for that passenger are closer to the real data.
 
 Step 5: We calculated the mean $M_t$ and standard error ${SD}_t$ of the set of relative errors $\Delta t'$. A mean value and variance closer to 0 indicate that the simulation system are closer to the real-world system.
 
\begin{figure}[!ht]
 	\centering
 	\includegraphics[width=0.5\textwidth]{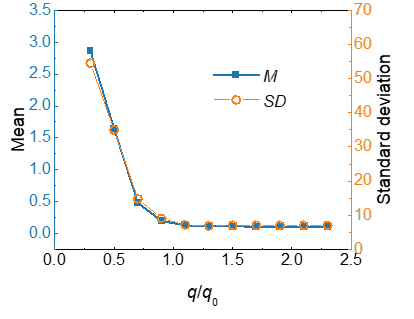}
 	\caption{Mean ($M$) and variance ($SD$) of the relative error set ($\Delta t'$) for the simulation system under different train capacities ($q_\mathrm{c}$). The blue curve represents the mean ($M$), while the orange curve represents the variance ($SD$). }
 	\label{fig:figS3}
 \end{figure}

 By comparing the mean and variance of the relative errors for different train capacities (Fig. \ref{fig:figS3}), we observed that as the train capacity increases, both $M_t$ and $SD$ decrease. When $q = 1.0$, $M_t$ and $S_t$ approach 0, indicating that the system corresponding to this set of train capacities is very close to the real-world system.
 
 In addition, we referred to the ``Urban Rail Transit Engineering Project Construction Standards (JB 104 2008)" for guidance. According to these standards, in the event of overcrowding, the seating capacity remains unchanged, while the standing capacity is calculated at 9 persons/m$^2$. Type B trains consist of 2 types: single-driver trains and non-driver trains. For single-driver trains, the overcrowding threshold is set at 230, while for non-driver trains, it is set at 250. Each train set includes a single-driver train at each end, with the remaining cars being non-driver trains. Notably, the subway lines studied (1, 2, 4, 5, 6, 7, 8, 10, and 13) utilize type B trains. However, lines 6 and 7 have 8-train sets, while the other lines have 6-train sets. Based on this information, we determined the maximum passenger capacity for each line and train set (refer to Table \ref{tab:table5}). The close alignment between these capacities and our model's parameter settings validates the effectiveness of our parameter adjustments.
 
 \section{Supplementary text}
 \subsection{Quantitative definition of the cluster}\label{sec:B.1}
 
 The number of stranded passengers at station $i$ at time $t$ is denoted as $S_i(t)$. The relation between 2 stranded stations, $i$ and $j$, can be classified into 2 categories:
 \begin{equation}
 	e_{ij}=
 	\begin{cases}
 		1& \text{ $ S_iS_j > 0$ and $T_{ij} = 1 $ } \\
 		0& \text{ else }.
 	\end{cases}
 	\label{eq:eqS2}
 \end{equation}
 
 The spatial correlation between station $i$ and station $j$ along the route is denoted by $T_{ij}$: $T_{ij} = 1$ if there is a station, $S_i\neq0$, that establishes connectivity between stations $i$ and $j$ on the route; otherwise, $T_{ij} = 0$. The propagation of stranding in the spatial domain and the formation of clusters composed of stranded stations are determined based on $e_{ij}$.
 
 \subsection{System I: Steady-state features}
 \subsubsection{Temporal variation of the average waiting time and the 3 order parameters}
\begin{figure}[!ht]
	\centering
	\includegraphics[width=0.7\textwidth]{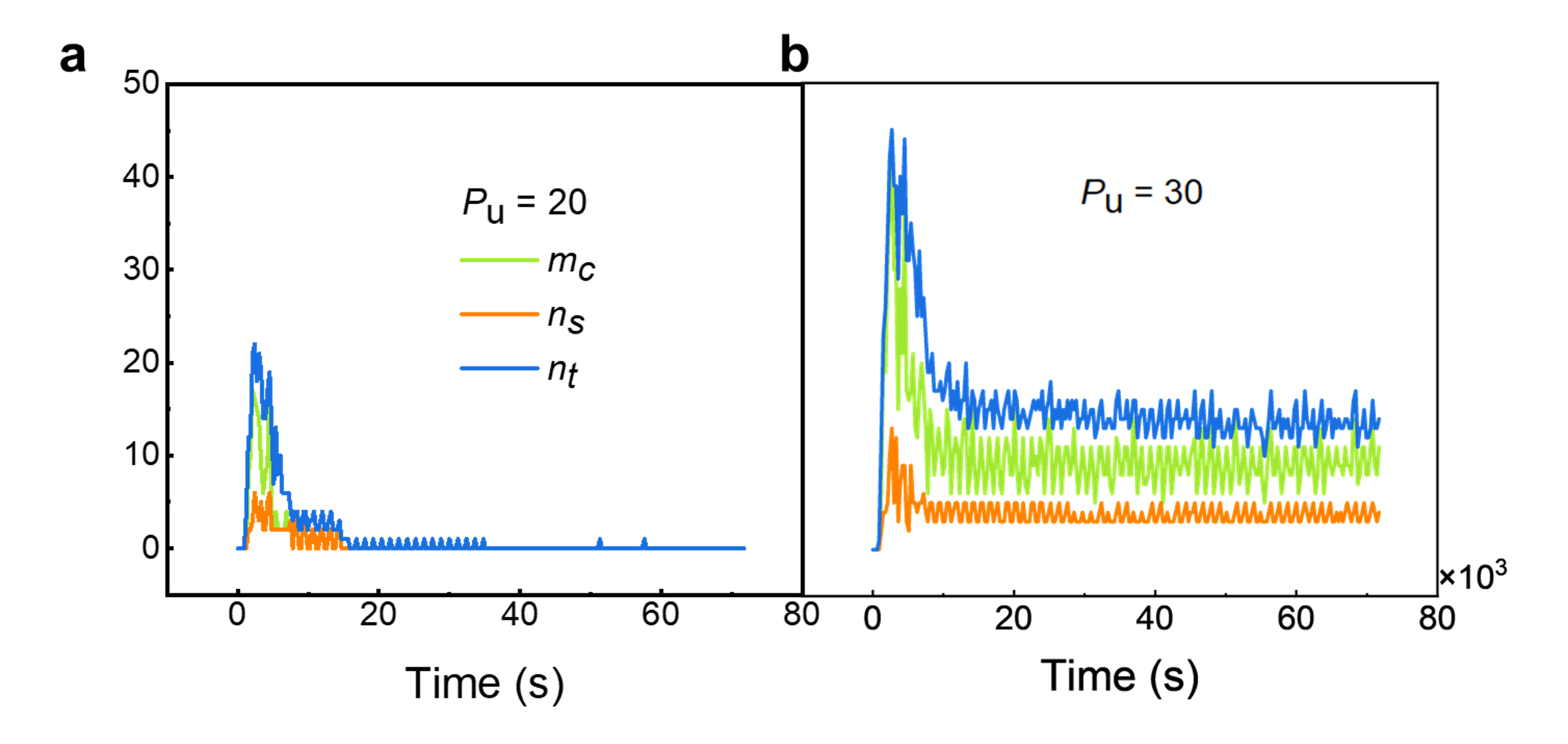}
	\caption{\textbf{a} The variation curves of $n_\mathrm{s}$, $m_\mathrm{c}$, and $n_\mathrm{t}$ in the system when $P_\mathrm{u} = 20$. \textbf{b} The variation curves of $n_\mathrm{s}$, $m_\mathrm{c}$, and $n_\mathrm{t}$ in the system when $P_\mathrm{u} = 30$. In fact, for all $P_\mathrm{u}$ values in the system, the 4 variables eventually reach a steady state. For clarity, only the changes in the number of clusters in the system for 2 $P_\mathrm{u}$ values are plotted.}
	\label{fig:figS4}
\end{figure}

 When the demand scale is relatively small, such as $P_\mathrm{u} = 20$, the evolution of the 3 order parameters over time is shown in Fig. \ref{fig:figS4}a. In the initial stages of the simulation, all vehicles in the system depart from the originating station, while passenger demands are randomly generated throughout the network. Due to insufficient service supply in the early stages, all 3 order parameters increase rapidly. However, after a relaxation time of $2.0\times10^4$, they reach a steady state. As the demand scale $P_\mathrm{u}$ increases (as shown in Fig. \ref{fig:figS4}b for $P_\mathrm{u} = 30$), the system reaches a steady state as long as the relaxation time is sufficiently long. In the steady state, all the order parameters significantly increase with the increasing demand scale $P_\mathrm{u}$. We use the average values of the 3 order parameters after the system reaches a steady state to characterize the state of the system under a given passenger demand scale $P_\mathrm{u}$ condition.

\subsubsection{Distribution evolution of station with stranded passengers in ideal model}

\begin{figure}[H]
	\centering
	\includegraphics[width=0.78\textwidth]{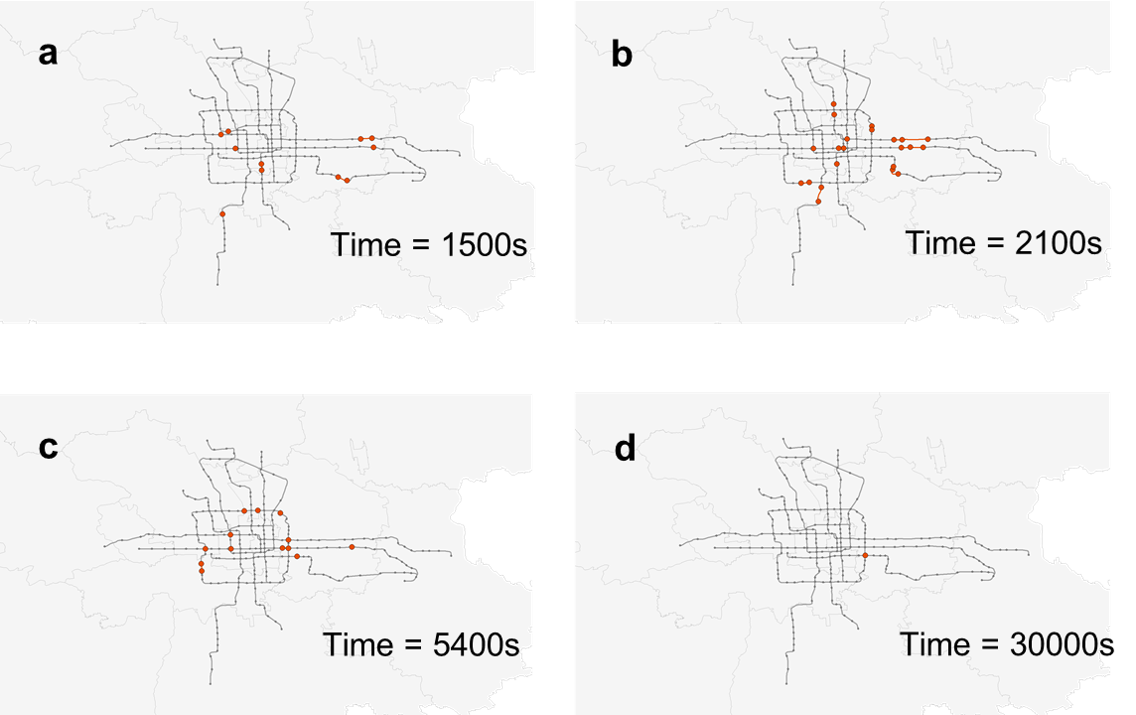}
	\caption{Evolution of station distribution for the ideal model with $P_\mathrm{u} = 20$.}
	\label{fig:figS5}
\end{figure}

\begin{figure}[H]
	\centering
	\includegraphics[width=0.78\textwidth]{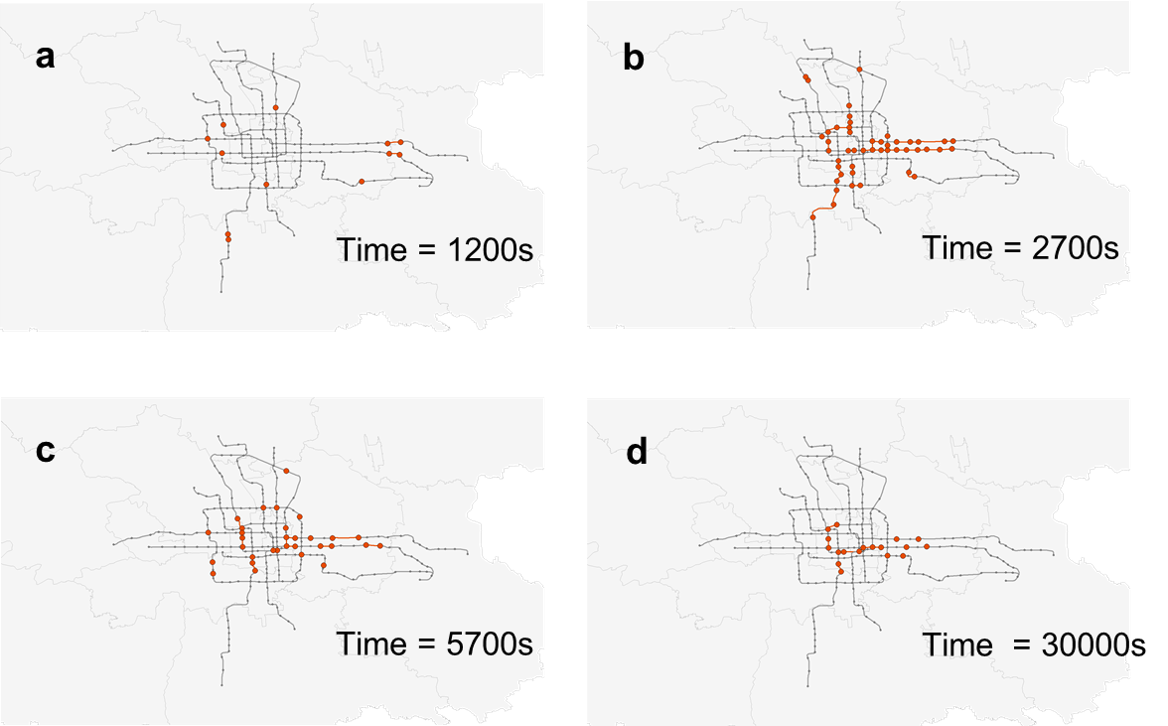}
	\caption{Evolution of station distribution for the ideal model with $P_\mathrm{u} = 30$.}
	\label{fig:figS6}
\end{figure}

\subsection{System II: Spatiotemporal mismatch between supply and demand}
\subsubsection{Tidal characteristics of demand and supply}

\begin{figure}[H]
	\centering
	\includegraphics[width=0.9\textwidth]{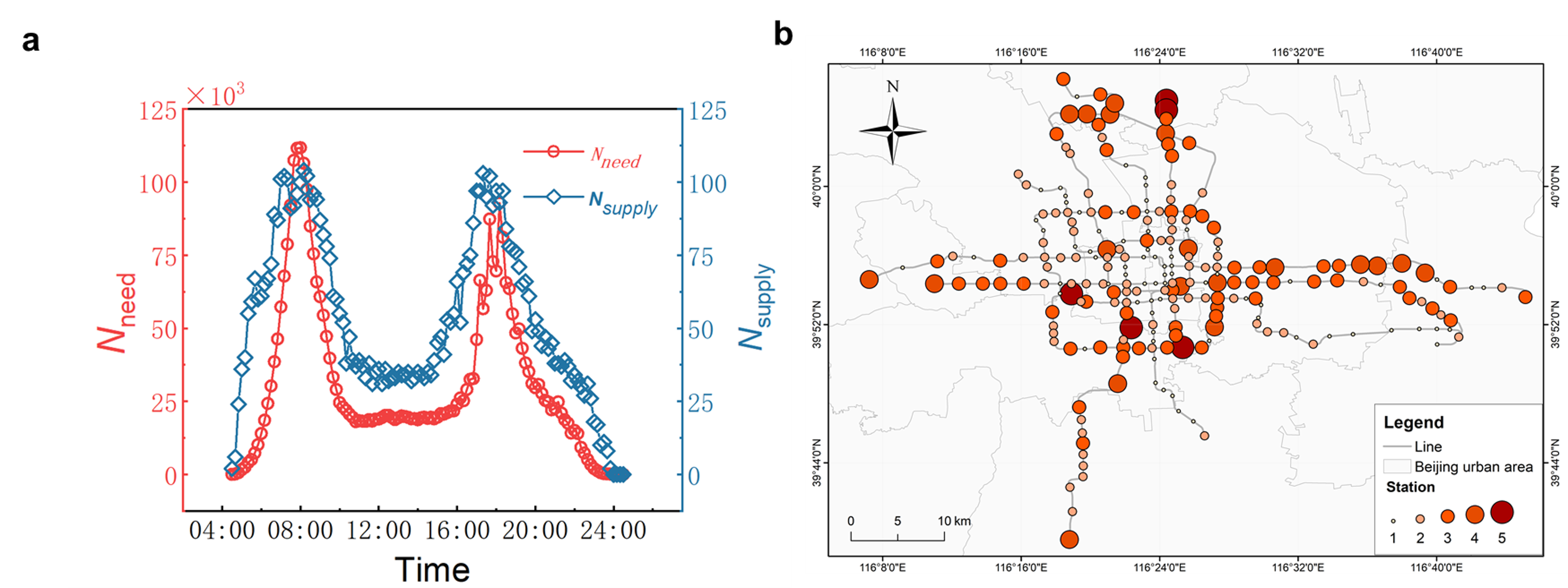}
	\caption{\textbf{a} Temporal distribution of demand and supply. Note: $N_\mathrm{supply}$ represents the total number of trains departing every 10 min in the model shown in Fig. \ref{fig:figS2}, and $N_\mathrm{need}$ represents the total number of passengers entering the station every 10 min in the model shown in Fig. \ref{fig:figS2}. \textbf{b} Spatial distribution of passenger demand between 6:00 and 13:00. The diameter of the loop line is proportional to the number of passengers entering the station at each location. Larger loop lines indicate higher passenger volumes, while smaller loop lines indicate lower volumes.}
	\label{fig:figS7}
\end{figure}

\subsubsection{Dynamics mechanism of cluster growth}
In the analysis of microstate, we employ the following statistical methods for the variables in the system.

(1) Number of stranded passengers: We introduce the variable $S_i^t$ to record the stranded status of passenger i. If a passenger fails to board the subsequent train after entering the station, they are considered stranded, and $S_i^t = 1$; otherwise, $S_i^t = 0$. We calculate the total number of stranded passengers at each station as the current station's count of stranded individuals. This count is updated every 10 minutes.

(2) Number of passengers entering each station: This data is obtained from the raw data, and we calculate the number of passengers entering each station at every time step.

(3) Number of passengers boarding: Since train arrivals at stations are discrete events, we adopt a method to account for variations caused by the presence or absence of trains during the data collection. If a train is currently in the passenger-loading state at a particular station, the number of boarding passengers is determined by summing the passenger count from that train's arrival time up to the current time. Otherwise, it is determined by the number of passengers carried by the last departing train from that station.

\subsubsection{Clusters and labeling}
\begin{figure}[H]
	\centering
	\includegraphics[width=0.95\textwidth]{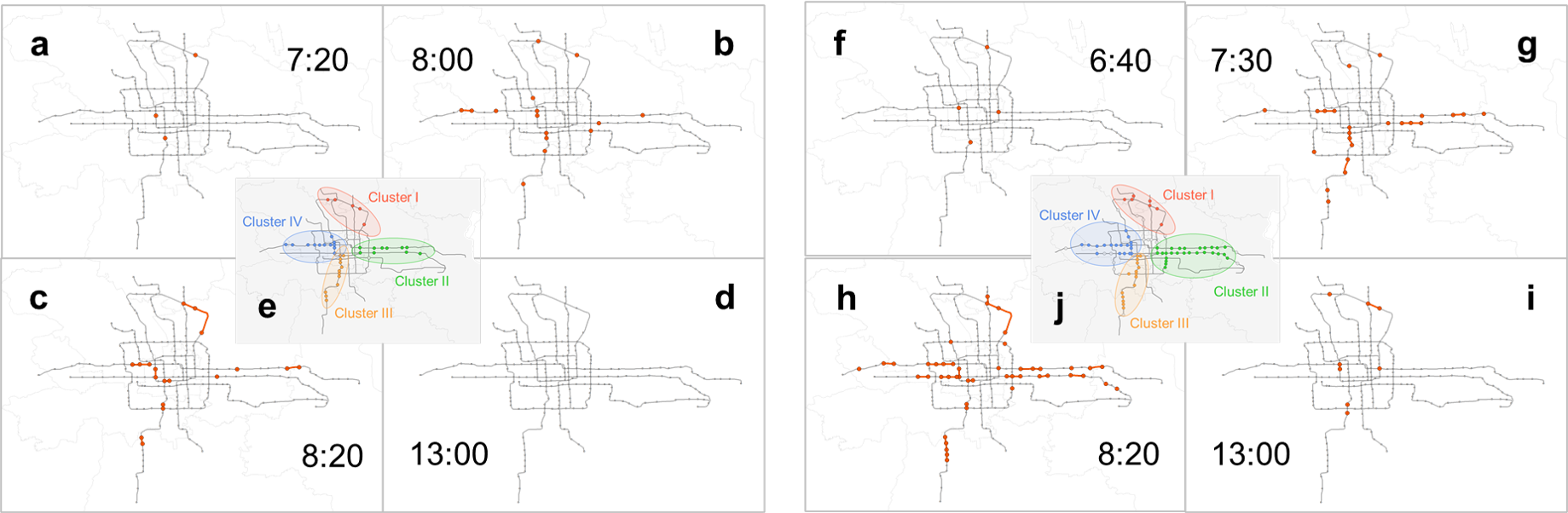}
	\caption{\textbf{a-d} Spatial distribution of critical stations in 4 clusters at 7:20, 8:00, 8:20, and 13:00 corresponding to $P_\mathrm{r} = 1.0$. \textbf{f-i} Spatial distribution of critical stations in 4 clusters at 6:40, 7:30, 8:20, and 13:00 in the network corresponding to $P_\mathrm{r} = 1.6$. \textbf{e} and \textbf{j} Clusters formed by critical stations that experienced stranded situations during the morning rush hours (6:30-13:00). These clusters consist of adjacent critical stations in both spatial and temporal dimensions. The stations included in Cluster \uppercase\expandafter{\romannumeral1}-\uppercase\expandafter{\romannumeral4} 4 are represented by green, blue, yellow, and purple colors, respectively.}
	\label{fig:figS8}
\end{figure}

During the evolution process of the real supply-demand system, if adjacent stations have experienced stranded passengers, they are considered to belong to the same cluster. During the morning rush hours on weekdays, 4 stranded station clusters have formed in 4 directions in Beijing (Fig. \ref{fig:figS8} presents critical stations for the 4 clusters for $P_\mathrm{r} = 1.0$ at $P_\mathrm{r} = 1.6$). The critical stations included in each cluster are listed in Table \ref{tab:table6} and Table \ref{tab:table7}.

\begin{table}[!ht]
	\caption{{Critical stations and critical stations' Label included in the 4 clusters formed during weekday morning rush hours at $P_\mathrm{r} = 1.0$}}
	\centering
	\begin{tabular}{cc}
		\toprule
		Cluster & Station(Label) \\
		\midrule
	 {Cluster1} &  {\makecell[c]{Huilongguan (230), Yuxin(200), Lishuiqiao(87),\\ Beiyuan(231), Wangjingxi(232)}} \\
		\midrule
		Cluster2 & Jintailu(119), Shilipu(120), Qingnianlu(121) \\
		\midrule
		 {Cluster3} &  {\makecell[c]{Hepingmen(46), Xuanwumen(47), Caishikou(67),\\ Taoranting(68), Beijing South Railway Station(69),\\ Majiapu(70), Jiaomenxi(71), Gongyi Xiqiao(72),\\ Xingong(73), Xihongmen(74), Gaomidianbei(75),\\ Gaomidiannan(76)}} \\
		\midrule
		 {Cluster4} &  {\makecell[c]{Fuxingmen(1), Fuchengmen(4), Chegongzhuang(50), \\Chegongzhuangxi(114),Baishiqiaonan(11),\\ Huayuanqiao(112), Cishousi(111), Haidian Wuluju(110)}} \\
		\bottomrule
	\end{tabular}
	\label{tab:table6}
\end{table}

\begin{table}[!ht]
	\caption{{Critical stations and critical stations' Label included in the 4 clusters formed during weekday morning rush hours at $P_\mathrm{r} = 1.6$}}
	\centering
	\begin{tabular}{cc}
		\toprule
		Cluster & Station(Label) \\
		\midrule
	 {Cluster1} &  {\makecell[c]{Huilongguan(230), Yuxin(200), LishuiQiao(87),\\ Beiyuan(231), Wangjingxi(232), Huoying(119),\\ Tiantongyuannan(86)}} \\ 
		\midrule
		 {Cluster2} &  {\makecell[c]{Jintailu(119), Shilipu(120), Qingnianlu(121),\\ Dalianpo(122), Huangqu(123), Changying(124),\\ Yong’anli(18), Guomao(19), Dawanglu(20),\\ Sihui(21), Sihuidong(22), Gaobeidian(23),\\ Communication University of China(24), Shuangqiao(25),\\ Guanzhuang(26), Baliqiao(27),\\ Tongzhou Beiyuan(28), Guoyuan(29)}} \\ 
		\midrule
		 {Cluster3} &  {\makecell[c]{Hepingmen(46), Xuanwumen(47), Caishikou(67),\\ Taoranting(68), Beijing South Railway Station(69),\\ Majiapu(70), Jiaomenxi(71), Gongyi Xiqiao(72),\\ Xingong(73), Xihongmen(74), Gaomidianbei(75),\\ Gaomidiannan(76), Caoqiao(159), Zaoyuan(77),\\ Qingyuanlu(78)}} \\ 
		\midrule
		 {Cluster4} &  {\makecell[c]{Fuxingmen(11), Fuchengmen(49),\\ Chegongzhuang(50), Chegongzhuangxi(114), \\Baishiqiaonan(113), Huayuanqiao(112),\\ Cishousi(111), Haidian Wuluju(110),\\ Xizhimen(51), Fuxingmen(10),\\ Muxidi(9), Military Museum(8)}} \\ 
		\bottomrule
	\end{tabular}
	\label{tab:table7}
\end{table}

\subsection{Validating the mechanism of cluster formation}

\subsubsection{Dynamics mechanism of cluster I growth}\label{sec:B.4.3}

\begin{figure}[!htb]
	\centering
	\includegraphics[width=0.5\textwidth]{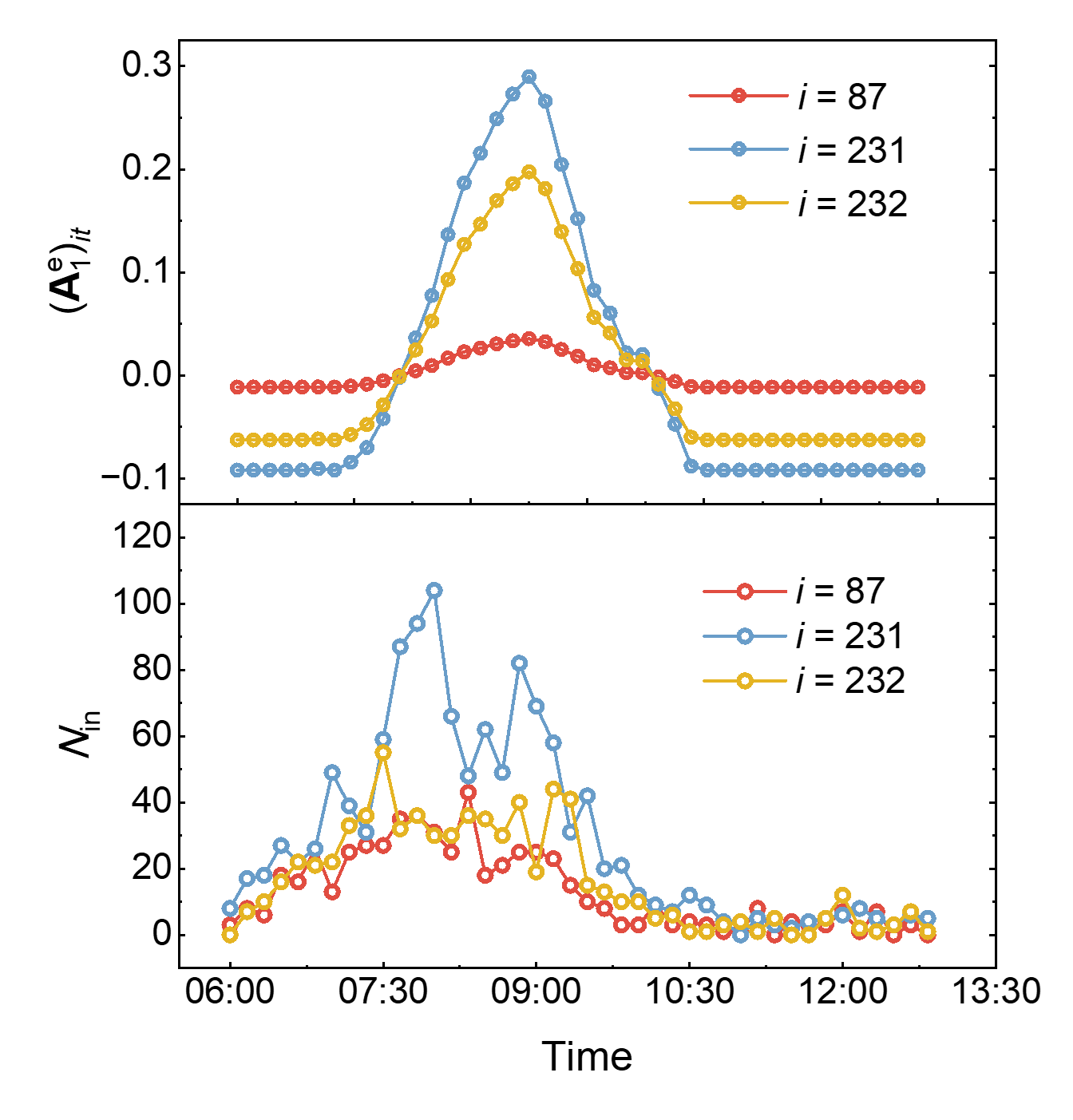}
	\caption{\textbf{a} Temporal evolution of the largest eigenstates in Cluster \uppercase\expandafter{\romannumeral1} for stations 87, 231, and 232 from 6:00 to 13:00. Stations 85, 86, 200, 88, and 89 do not exhibit significant variations in their eigenstates and are therefore not shown. \textbf{b} Variation in the number of entering passengers for stations 87, 231, and 232 from 6:00 to 13:00.
	}
	\label{fig:fig6}
\end{figure}

\textbf{Local pattern}: The largest eigenvalue, responsible for 85.9\% of the total contribution, exhibits significant spatiotemporal variations at 3 distinct stations: $i = 87$, 231, and 232. The temporal evolution of ${{(A}_1^\mathrm{e})}_{it}$ at these stations is depicted in Fig. \ref{fig:fig6}a, demonstrating a characteristic pattern with a single peak during the early peak period, reaching its maximum around 8:50, and returning to the initial state by approximately 10:40. Station 231 exhibits the most prominent amplitude, while station 87 shows a more gradual variation. These observed patterns coincide with the temporal changes in the number of passengers entering the corresponding stations (i.e., demand scale), suggesting that this stranded pattern arises from an increase in the same passenger demand. Thus, we term this phenomenon the Local pattern. Consequently, the primary driving factor behind the increase in stranded passengers at these stations is the mismatch between supply and demand caused by localized demand growth.

\begin{figure}[H]
	\centering
	\includegraphics[width=1.0\textwidth]{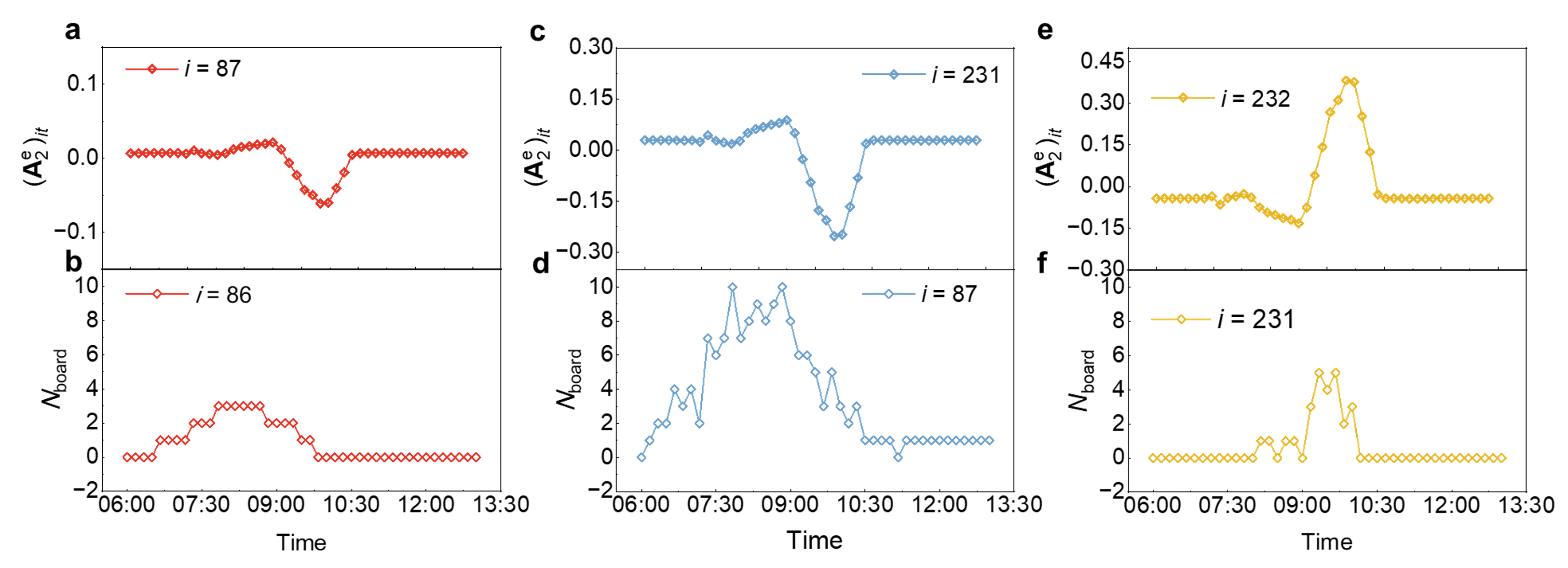}
	\caption{Variation of the second-largest eigenstates and passenger demand in Cluster \uppercase\expandafter{\romannumeral1}. \textbf{a, c, and e} depict the temporal evolution of the second-largest eigenstates for stations 87, 231, and 232, respectively, from 6:00 to 13:00. \textbf{b, d, and f} represent the changes in the number of boarding passengers at upstream stations 86, 87, and 231, respectively. The eigenstates of other stations within Cluster \uppercase\expandafter{\romannumeral1} exhibit minimal variations.
	}
	\label{fig:fig7}
\end{figure}

\textbf{Nonlocal pattern}: The second-largest eigenvalue accounts for 13.4\% of the total variance. Figs. \ref{fig:fig7}a, c and e depict the temporal evolution of ${{(A}_2^\mathrm{e})}_{it}$ for stations 87, 231, and 232, respectively. In contrast to the characteristic single peak of Local pattern, this pattern exhibits alternating positive and negative variations, resembling the derivative of a single peak function. By comparing the morning rush hour stranded passenger counts at upstream stations $i = 86$, 87, and 231 (Figs. \ref{fig:fig7}b, d, f) with ${{(A}_2^\mathrm{e})}_{it}$, it can be inferred that the mechanism of stranded conditions involves propagation. This suggests that the variations in this pattern are driven by an increase in boarding passengers at upstream stations during rush hour periods. The stranded conditions at upstream stations result in a reduction of service supply downstream, leading to an increase in stranded passengers at downstream stations. We named this phenomenon the Nonlocal pattern. Specifically, between 7:50 and 9:00, the increase in boarding passengers at station 86 leads to overcrowding, and this overcrowded condition does not improve at the nearest downstream station, 87. This interdependent relationship between adjacent stations facilitates the propagation of stranded conditions along the line direction. This effect is a secondary consequence of the Local pattern, resulting in a lower contribution from the Nonlocal pattern. Similarly, the increase in boarding passengers at station 231 exacerbates the stranded conditions downstream by reducing supply. During this period, both station 87 and station 231 experience stranded conditions, which have a suppressing effect on downstream station 232. Between 9:00 and 10:40, as the number of boarding passengers at station 231 increases, the stranded conditions at downstream station 232 worsen. Simultaneously, the decrease in boarding passengers at station 86 gradually reduces the occupation of supply at downstream stations, thereby alleviating the stranded conditions. Additionally, the impact of station 87 on the stranded conditions at station 231 starts to weaken. This mode reflects the propagation of stranded conditions and explains how the traffic supply at upstream stations influences the stranded conditions at downstream stations.

\begin{figure}[!ht]
	\centering
	\includegraphics[width=0.55\textwidth]{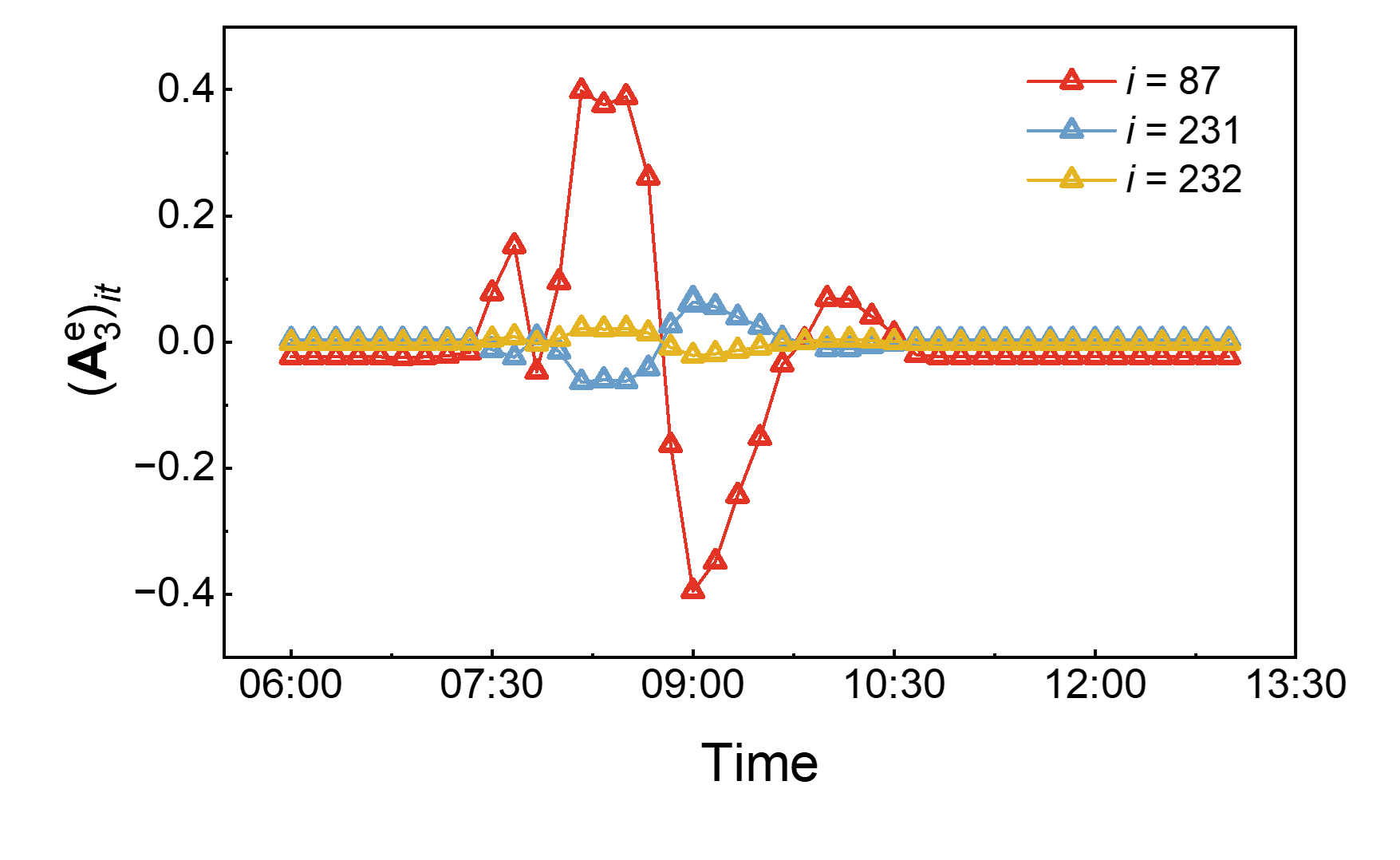}
	\caption{Temporal evolution of the third-largest eigenstates for stations 87, 231, and 232 from 6:00 to 13:00.
	}
	\label{fig:fig8}
\end{figure}

\textbf{Topological pattern}: The third-largest eigenvalue corresponds to a pattern contribution of 1.7\%. Fig. 5 illustrates the temporal evolution of ${{(A}_3^\mathrm{e})}_{it}$. This pattern unveils the influence of the infrastructure's topological structure on passenger stranding, which we refer to as the Topological pattern. The distinguishing characteristic of this pattern is the significant variation observed at a specific station, designated as Local pattern, as depicted in Fig.~\ref{fig:fig8} for station 87. Unlike the other 2 stations, station 87 functions as a transfer station, with multiple upstream stations contributing to the propagation of stranding and multiple downstream stations contributing to stranding alleviation, while stations 231 and 232 serve as non-transfer stations. Consistent with the findings from steady-state condition calculations under spatially and temporally homogeneous demand conditions discussed in the previous section, stations with high network betweenness centrality at transfer points play a pivotal role in PPS occurring. Under steady-state conditions, only the topological structure of the network influences PPS. In this system characterized by spatially and temporally heterogeneous supply and demand, the contribution of the topological factor ranks third.

\subsubsection{Dynamics mechanism of II-IV growth}
\begin{figure}[H]
	\centering
	\includegraphics[width=1.0\textwidth]{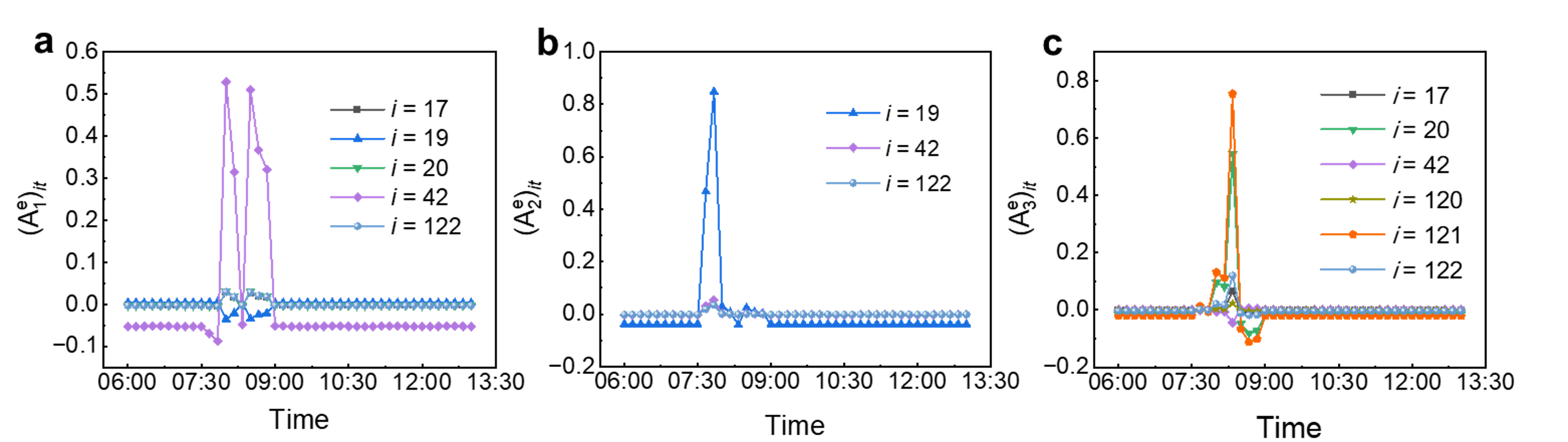}
	\caption{\textbf{a-c} Topological Pattern, Nonlocal Pattern, and Local Mode of Cluster \uppercase\expandafter{\romannumeral2}.}
	\label{fig:figS9}
\end{figure}

\begin{figure}[!ht]
	\centering
	\includegraphics[width=1.0\textwidth]{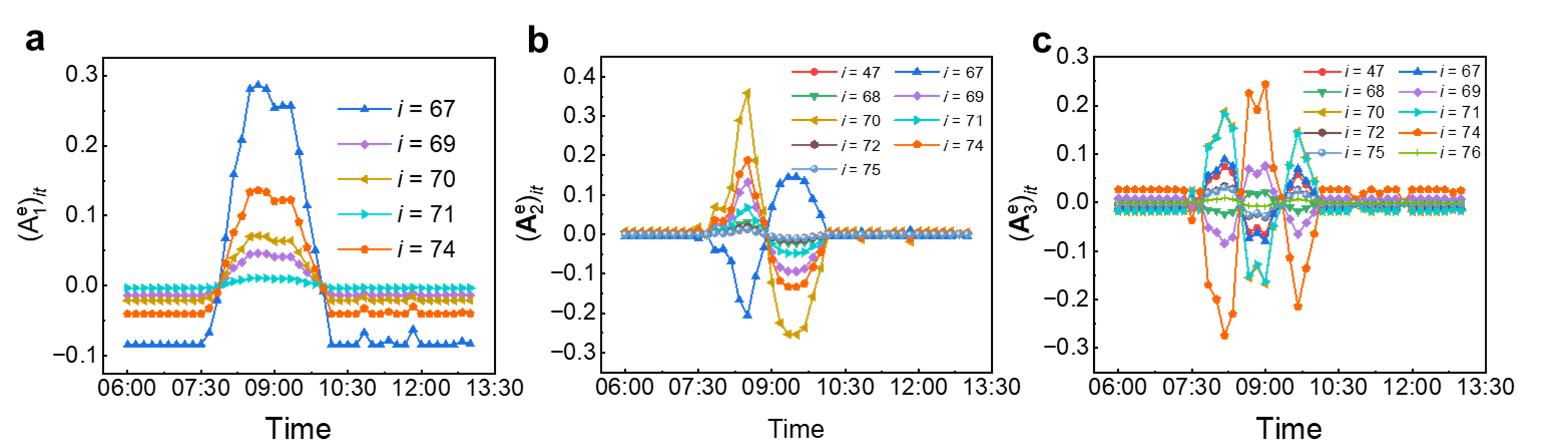}
	\caption{\textbf{a-c} Topological Pattern, Nonlocal Pattern, and Local Mode of Cluster \uppercase\expandafter{\romannumeral3}.}
	\label{fig:figS10}
\end{figure}

\begin{figure}[H]
	\centering
	\includegraphics[width=1.0\textwidth]{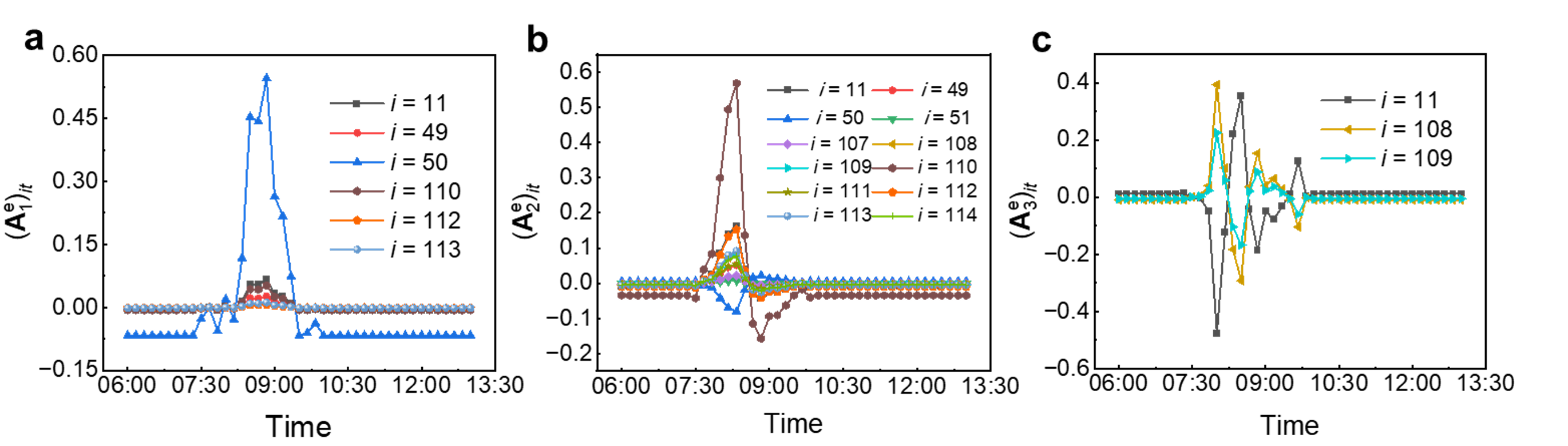}
	\caption{\textbf{a-c} Topological pattern, Nonlocal pattern, and Local pattern of Cluster \uppercase\expandafter{\romannumeral4}.}
	\label{fig:figS11}
\end{figure}

\subsubsection{Train control scheme}\label{sec:B.4.2}
\begin{figure}[!ht]
	\centering
	\includegraphics[width=0.5\textwidth]{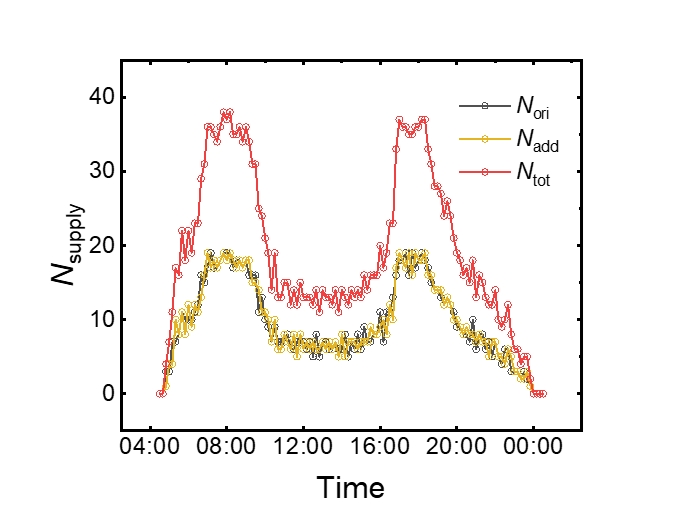}
	\caption{Changes in the Number of Train Services for the Original Timetable ($N_\mathrm{ori}$), Additional Timetable ($N_\mathrm{add}$), and New Timetable ($N_\mathrm{tot}$) for Lines 5 and 13.}
	\label{fig:figS12}
\end{figure}

We maintain the spatial distribution of train departures while adjusting the frequency of train departures on Lines 5 and 13 to modify the supply scale. Initially, we preserve the spatial distribution of train departures according to the original timetable for Lines 5 and 13. We double the train departure density, resulting in a total number of train services denoted as $N_\mathrm{tot}$. The additional trains in the expanded timetable are scheduled to depart 2 min later than the original timetable (Fig. \ref{fig:figS12}). Since the time delay is relatively short, the new timetable preserves the spatiotemporal characteristics of the original one. Based on the new train timetable, we randomly select the required number of train services according to passenger demand to form the actual subway timetable. Let $N_\mathrm{ori}$ denote the total number of train services in the original timetable. We define the adjusted supply, $P_\mathrm{s}$, as $N_\mathrm{tot}/ N_\mathrm{ori}$.

\subsubsection{Changes in the contributions of different supply scale patterns}\label{sec:B.4.4}

\begin{figure}[H]
	\centering    
	\includegraphics[width=1.0\textwidth]{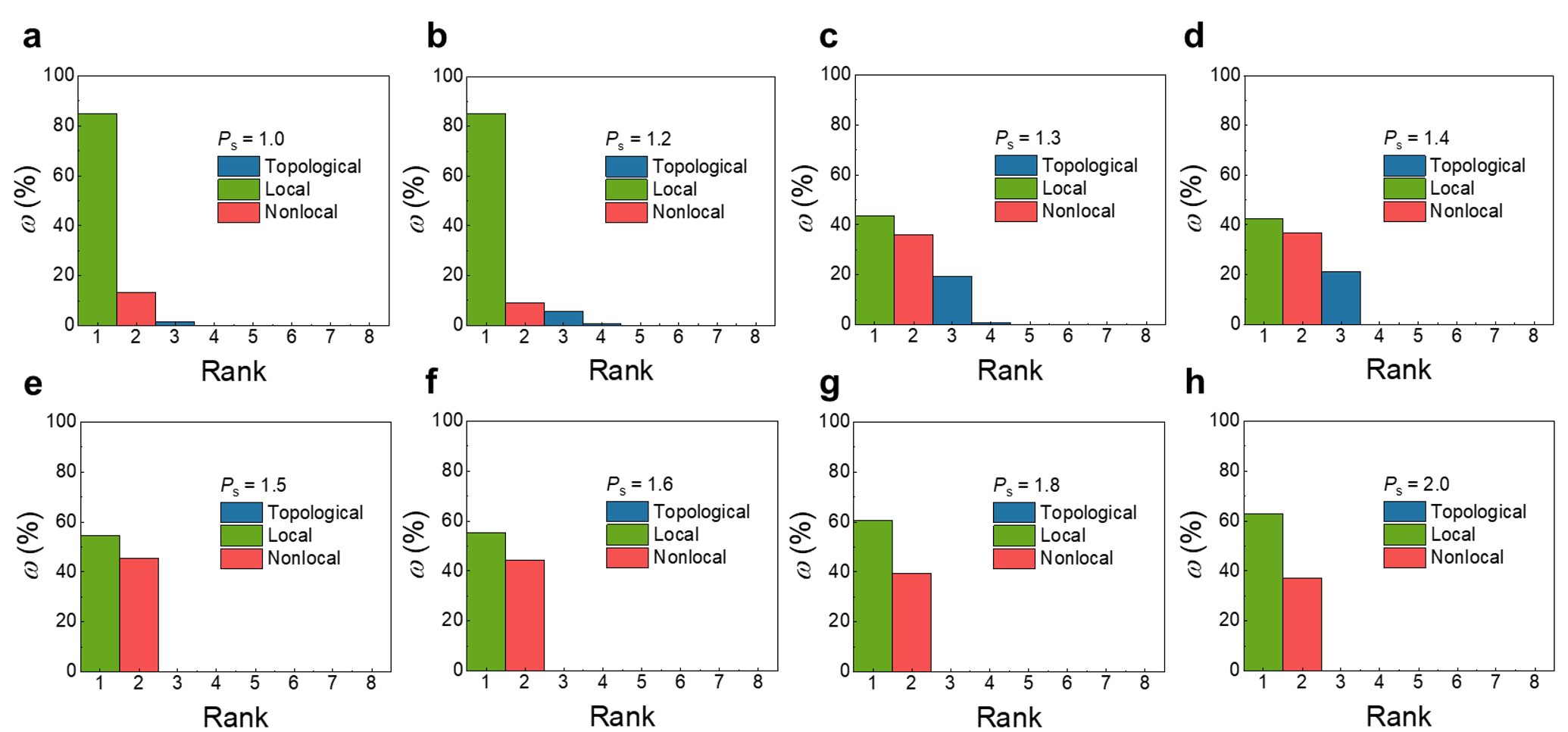}
	\caption{Proportional contributions of Local (green), Nonlocal (red), and Topological (blue) patterns for different $P_\mathrm{s}$. \textbf{a-h} respectively represent contribution ratios of 3 patterns for$P_\mathrm{s} = 1.0, 1.2, 1.3, 1.4, 1.5, 1.6, 1.8, 2.0$.}
	\label{fig:figS13}
\end{figure}
% \end{document}


\begin{thebibliography}{111}
	\bibitem{1}Peng, Y.-T.,   Li, Z.-C. \&  Choi, K. Transit-oriented development in an urban rail transportation corridor. \emph{Transport. Res. B-Meth.} 	\textbf{103}, 269-290 (2017).
	
	\bibitem{2}  Liu, X. et al.  	Network resilience. \emph{Phys. Rep.} \textbf{971}, 1-108 (2022).
	
	\bibitem{3} Holling, C. S. Resilience and stability of ecological systems. \emph{Annu. Rev. Ecol. Evol. Syst.} \textbf{4}, 1-23 (1973).
	
	\bibitem{4} National Academy of Engineering \&   Schulze P. \emph{Engineering Within Ecological Constraints} (National Academy Press, New York,  1996).
	
	
	
	
	\bibitem{5} Li, D. et al. Percolation transition in dynamical traffic network with evolving critical bottlenecks. \emph{Proc. Natl. Acad. Sci. U. S. A.} \textbf{112}, 669-672 (2015).
	
	\bibitem{6}  Peng, H. et al. Spatial temporal incidence dynamic
	graph neural networks for traffic flow forecasting. \emph{Inf. Sci.} \textbf{521}, 277-290 (2020).
	
	\bibitem{7}  Lighthill, M. J. \&   Whitham, G. B.  On kinematic waves I. Flood movement in long rivers. \emph{Proc. R. Soc. Lond. A}  \textbf{229}, 281-316 (1955).
	
	\bibitem{8}  Prigogine, I., Herman,  R. \&   Chaiken, J. Kinetic theory of vehicular traffic. \emph{Phys. Today}  \textbf{25} 56–57 (1972).
	
	
	
	\bibitem{9}Newell, G. F.  A simplified theory of kinematic waves in highway traffic, part i: General theory. \emph{Transport. Res. B-Meth.} \textbf{27}, 281-287 (1993).
	
	\bibitem{10} Band\={o}, M.,  Hasebe, K.,   Nakayama, A.,  Shibata, A. \&  Sugiyama, Y. Dynamical model of traffic congestion and numerical simulation. \emph{Phys. Rev. E: Stat. Phys., Plasmas, Fluids, Relat. Interdiscip. Top.} \textbf{51}, 1035-1042 (1995).
	
	\bibitem{11}  Treiber, M.,  Hennecke, A. \&  Helbing, D. Congested traffic states in empirical observations and microscopic simulations. \emph{Phys. Rev. E: Stat. Phys., Plasmas, Fluids, Relat. Interdiscip. Top.} \textbf{62}, 1805-1824 (2000).
	
	\bibitem{12}  Nagel, K. \&   Schreckenberg, M. A cellular automaton model for freeway traffic. 	\emph{J. Phys. I} \textbf{2}, 2221-2229 (1992).
	
	
	\bibitem{13} Zhang, X., Miller-Hooks, E. \&  Denny, K. Assessing the role of network topology in transportation	network resilience. \emph{J. Transp. Geogr.} \textbf{46}, 35-45 (2015).
	
	
	\bibitem{14}   Lu, Q.-C. Modeling network resilience of rail transit under operational incidents. \emph{Transp. Res. Part A Policy Pract.} \textbf{117}, 227-237  (2018).
	
	\bibitem{15}  Jin, J. G.,  Tang,  L. C.,  Sun, L. \&   Lee, D.-H. Enhancing metro network resilience	via localized integration with bus services. \emph{Transport. Res.  E-Log.} \textbf{63}, 17-30 (2014).
	
	\bibitem{16}  Montis, A. D.,    Barthelemy, M.,    Chessa, A. \&   Vespignani, A. The structure of
	inter-urban traffic: a weighted network approach.   \emph{Environ. Plann. B Plann. Des.} \textbf{34}, 905-924 (2007).

	
	\bibitem{17} Izawa, M. M.,   Oliveira, F. A.,    Cajueiro, D. O. \& Mello B. A.
	Pendular behavior of public transport networks.  \emph{Phys. Rev. E} \textbf{96},  012309 (2017).
	
	\bibitem{18}  Li, J. \&   Deepak, F. L. In situ kinetic observations on crystal nucleation and growth. \emph{Chem. Rev.}    \textbf{122},   16911-16982
(2022).
	
	\bibitem{19} Schell, C. J. et al.  The ecological and evolutionary
	consequences of systemic racism in urban environments. \emph{Science}  \textbf{369}, eaay4497  (2020).
	
	
	
	\bibitem{20}Gibbs, J. W. \emph{Elementary Principles in Statistical Mechanics}  (Cambridge University Press, Cambridge, 1902).
	
	\bibitem{21}   Hu, G.,  Liu, T.,  Liu,  M.,  Chen, W. \&   Chen, X. Condensation of eigen microstate in statistical ensemble and phase transition. \emph{Sci. China Phys. Mech. Astron.} \textbf{62}, 1-8 (2018).
	
	\bibitem{22}  Wang, N.-N.,   Qiu, S.,  Zhong, X. \&   Di, Z. Epidemic thresholds identification of	susceptible-infected-recovered model based on the eigen microstate. \emph{Appl. Math. Comput.} \textbf{449}, 127924 
	(2023).
	
	\bibitem{23}  Geng, Z. et al.   Networksynchronization	analysis reveals the weakening tropical circulations. \emph{Geophys. Res. Lett.} \textbf{48},  e2021GL093582 (2021).
	
	
	\bibitem{24}  Liu, T. et al.  Renormalization	group theory of eigen microstates. \emph{Chinese Phys. Lett.} \textbf{39}, 080503 (2022).
	
	\bibitem{25}  Fan, J. et al.   Network-based approach and climate change benefits for forecasting the amount of indian monsoon rainfall.  \emph{J. Clim.}  \textbf{35}, 10090-1020 (2020).
	
	\bibitem{26}  Yang, Q.,    Yang, B.,   Qiao, Z.,   Tang, M. \&   Gao, F. Impact of possible random factors on
	queue behaviors of passengers and taxis at taxi stand of transport hubs. \emph{Phys. A} \textbf{580}, 126131  (2021).
	
	
	\bibitem{27}    Yan, Y.,  Meng, Q.,  Wang, S. \&   Guo, X. Robust optimization model of schedule
	design for a fixed bus route.  \emph{Transp. Res. Part C Emerg. Technol.} \textbf{25}, 113-121 (2012).
	
	\bibitem{28}   Nagatani, T. Bunching transition in a time-headway model of a bus route. \emph{Phys. Rev. E: Stat., Nonlinear, Soft Matter Phys.} \textbf{63}, 036115 (2001).
	
	\bibitem{29}  Markos P. \&  Dentsoras, A. J. An integrated mathematical method for traffic analysis of elevator systems. \emph{Appl. Math. Model.} \textbf{105}, 50-80 (2021).
	
		\bibitem{30}   Nagatani, T. Complex motion induced by elevator choice in peak traffic. \emph{Phys. A} \textbf{436}, 159-169 (2015).
		
		\bibitem{31}   Nagatani, T. Dynamical transitions in peak elevator traffic. \emph{Phys. A} \textbf{333}, 441-452  (2004).
		
	\bibitem{32}   Luh, P. B.,   Xiong, B. \&   Chang, S.-C. Group elevator scheduling with advance information for
	normal and emergency modes.  \emph{IEEE Trans. Autom. Sci. Eng.} \textbf{5}, 245-258 
	(2008).
	
	\bibitem{33}   Liang, C.,  Hu, X.,  Shi, L.,   Fu, H. \&   Xu, D. Joint dispatch of shipment equipment
	considering underground container logistics. \emph{Comput. Ind. Eng.} \textbf{165}, 107874  (2022).
	
	\bibitem{34}   Wong, E. Y. C.,  Tai, A. H. \&   So, S. Container drayage modelling with graph
	theory-based road connectivity assessment for sustainable freight transportation in new development
	area. \emph{Comput. Ind. Eng.} \textbf{149}, 106810  (2020).
	
	\bibitem{35}   Li, C.,   Zhang, Y.,  Su,  X. \&   Wang, X. An improved optimization algorithm for	aeronautical maintenance and repair task scheduling problem. \emph{Mathematics} \textbf{10}, 3777  (2022).
	
	\bibitem{36}  Su,  X.,  Cui,  R.,  Li, C.,   Han, W. \&   Liu, J. A heuristic solution framework for the
	resource-constrained multi-aircraft scheduling problem with transfer of resources and aircraft. \emph{Expert
	Syst. Appl.} \textbf{228}, 120430  (2023).
	
	\bibitem{37}   Liao,  S. S.,   Moore, T. P. \&     Mackel, A. G. A transportation and logistic support
	model for aircraft aboard navy carriers. \emph{Transp. Res. Part A Policy Pract.} \textbf{26}, 231-245
	(1992).
	\bibitem{38}   Strang, G. \emph{Introduction to Linear Algebra, Sixth Edition} (Wellesley-Cambridge Press, Cambridge, 2024).
	
	\bibitem{39}  Sun, Y. et al. Eigen microstates and their evolutions in complex systems. \emph{Commun. Theor. Phys.} \textbf{73}, 065603 (2021).
    
     \bibitem{40}   Hu, G. , Liu, T. , Liu, M. , Chen, W. , \& Chen, X. Condensation of eigen microstate in statistical ensemble and phase transition. \emph{Sci. China Phys.} \textbf{62}, 990511 (2019)

\bibitem{41}   Tay, C. C.,    Sakidin, H.,      Jamaludin, I. W.,  Razak, N. A.  \&    Apandi, N. I. A. \emph{Introduction to Linear Algebra}  (Penerbit UTeM Press, Penerbit, 2010). 
	\end{thebibliography}
\end{document}